\documentclass[12pt]{article}
\usepackage{geometry}                
\usepackage{color}
\usepackage[pdftex]{graphicx}
\usepackage{subfigure}
\usepackage{wrapfig}
\usepackage{amssymb}
\usepackage{epstopdf}
\usepackage{float}
\usepackage{mathtools}
\usepackage{caption}
\usepackage{multirow}
\usepackage{hyperref}

\graphicspath{{./MC_images/}}
\newcommand{\be}{\begin{equation}}       
\newcommand{\ee}{\end{equation}}       
\newcommand{\bear}{\begin{eqnarray}}       
\newcommand{\eear}{\end{eqnarray}}       
\newcommand{\ba}{\begin{array}}       
\newcommand{\ea}{\end{array}}

\newskip\humongous \humongous=0pt plus 1000pt minus 1000pt

\newif\ifdtup

\def\oldreffmt#1{\rlap{[#1]} \hbox to 2\parindent{}}

\def\figfmt#1{\rlap{Figure {#1}} \hbox to 1in{}}       
       
\newskip\humongous \humongous=0pt plus 1000pt minus 1000pt

\newif\ifdtup

\def\oldreffmt#1{\rlap{[#1]} \hbox to 2\parindent{}}

\def\figfmt#1{\rlap{Figure {#1}} \hbox to 1in{}}  
  
\def\ie{\hbox{\it i.e.}{}}        
\def\eg{\hbox{\it e.g.}{}}        
\def\etal{\hbox{\it et al.}}  
  
\def\beq{\begin{equation}}  
\def\eeq{\end{equation}}  
\def\bea{\begin{eqnarray}}  
\def\eea{\end{eqnarray}}

\def\bq{\begin{quote}}  
\def\eq{\end{quote}}

\def \lta {\mathrel{\vcenter  
     {\hbox{$<$}\nointerlineskip\hbox{$\sim$}}}}  
   
\def \etal {{\it et al.}\ } 
\def\mh{m_{h}}

\def\gmh{\Gamma_h}
\def\mm{\mu^+ \mu^-}

\def\fbi{\,{\rm fb}^{-1}}
\def\fb{\,{\rm fb}}
\def\pbi{\,{\rm pb}^{-1}}
\def\pb{\,{\rm pb}}
\def\br{\,{\rm Br}}
\def\cm2s{\,{\rm cm}^{-2} {\rm s}^{-1}}

\newcommand{\mev}{\,\text{MeV}}
\newcommand{\gev}{\,\text{GeV}}
\newcommand{\tev}{\,\text{TeV}}

\relax  

\newdimen\tdim  
\tdim=\unitlength  
\def\bar{\overline}

\newskip\humongous \humongous=0pt plus 1000pt minus 1000pt

\newif\ifdtup

\def\oldreffmt#1{\rlap{[#1]} \hbox to 2\parindent{}}

\def\figfmt#1{\rlap{Figure {#1}} \hbox to 1in{}}  
  
\def\ie{\hbox{\it i.e.}{}}        
\def\eg{\hbox{\it e.g.}{}}        
\def\etal{\hbox{\it et al.}}  
  
\def\beq{\begin{equation}}  
\def\eeq{\end{equation}}  
\def\bea{\begin{eqnarray}}  
\def\eea{\end{eqnarray}}

\def\bq{\begin{quote}}  
\def\eq{\end{quote}}

\def \lta {\mathrel{\vcenter  
     {\hbox{$<$}\nointerlineskip\hbox{$\sim$}}}}  
   
\def \etal {{\it et al.}\ }  
\relax  

\newdimen\tdim  
\tdim=\unitlength  
\def\bar{\overline}

\def\cm{c_-}

\def\h{h}
\def\mh{m_{\h}}

\def\anti{\overline}

\def\mev{~{\rm MeV}}
\def\gev{~{\rm GeV}}
\def\tev{~{\rm TeV}}

\def\fbi{~{\rm fb}^{-1}}

\def\beq{\begin{equation}}
\def\eeq{\end{equation}}
\def\bea{\begin{eqnarray}}
\def\eea{\end{eqnarray}}

\begin{document}
\baselineskip=18pt \pagestyle{plain} \setcounter{page}{1}

\begin{flushright}       
UCLA Physics Preprint-100\\
FERMILAB-CONF-13-245-T \\ [1mm]       
August 8, 2013       
\end{flushright}

\vspace*{1.2cm}

\begin{center}
{\Large \bf
Muon Collider Higgs Factory\\ 
for \\
 Snowmass 2013
} \\ [9mm]

{\normalsize 
  Yuri ~Alexahin$^{(1)}$,
  Charles~M.~Ankenbrandt$^{(2)}$,
  David~B.~Cline$^{(3)}$, 
  Alexander~Conway$^{(4)}$, 
  Mary~Anne~Cummings$^{(2)}$, 
  Vito~Di~Benedetto$^{(1)}$, 
  Estia~Eichten$^{(1)}$, 
  Corrado~Gatto$^{(5)}$,
  Benjamin~Grinstein$^{(6)}$,
  Jack~Gunion$^{(7)}$,  
  Tao~Han$^{(8)}$, 
  Gail~Hanson $^{(9)}$,  
  Christopher~T.~Hill$^{(1)}$, 
  Fedor~Ignatov$^{(10)}$,
  Rolland~P.~Johnson$^{(2)}$,
  Valeri~Lebedev$^{(1)}$,
  Ron~Lipton$^{(1)}$, 
  Zhen~Liu $^{(8)}$, 
  Tom~Markiewicz$^{(11)}$,
   Anna~Mazzacane$^{(1)}$,
   Nikolai~Mokhov$^{(1)}$,
   Sergei Nagaitsev $^{(1)}$, 
   David~Neuffer$^{(1)}$, 
   Mark~Palmer$^{(1)}$, 
   Milind~V.~Purohit$^{(12)}$,
   Rajendran Raja$^{(1)}$,
   Sergei~Striganov$^{(1)}$,
   Don~Summers$^{(14)}$,  
   Nikolai~Terentiev$^{(15)}$,  
   Hans~Wenzel$^{(1)}$
                             } \\ [1cm]

{\small (1) {\it Fermilab, P.O. Box 500, Batavia, Illinois, 60510; } \\ }
{\small (2) {\it Muons Inc., Batavia, Illinois, 60510 } \\ }
{\small (3) {\it University of California, Los Angeles, California; }\\ }
{\small (4) {\it University of Chicago, Chicago, Illinois; }\\ }
{\small (5) {\it INFN Naples, Universita Degli Studi di Napoli Federico II, Italia;}\\ }
{\small (6) {\it University of California, San Diego, California; }\\ }
{\small (7) {\it University of California, Davis, California; }\\ }
{\small (8) {\it University of Pittsburgh, Pittsburgh, Pennsylvania;}\\ }
{\small (9) {\it University of California, Riverside, California; }\\ }
{\small (10) {\it Budker Institute of Nuclear Physics, Russia;}\\ }
{\small (11) {\it SLAC National Accelerator Laboratory, Menlo Park, California;}\\}
{\small (12) {\it University of  South Carolina,  Columbia, South Carolina; }\\ }
{\small (13) {\it University of California, Irvine, California; }\\ }
{\small (14) {\it University of Mississippi, Oxford, Miss.; }\\ }
{\small (15) {\it Carnegie Mellon University, Pittsburgh, Pennsylvania;}\\ }

\end{center}

\vspace*{0.1cm}

\newpage
                         
\section*{Executive Summary}

We propose the construction of a 
compact Muon Collider $s$-channel Higgs Factory. 
A Muon Collider Higgs Factory is part of an evolutionary program 
beginning with R\&D on Muon Cooling with a possible neutrino factory such as $\nu$STORM, the construction of Project-X with a rich program of precision physics addressing the $\sim 100 \tev$ scale, potentially leading ultimately to the construction of an energy frontier Muon Collider with $\mu^+$ and $\mu^-$ colliding up to $\sim 10.0 \tev$  center-of-mass energy.  The Muon Collider Higgs Factory would utilize an intense proton beam from Project-X.

The Higgs boson is a particle of fundamental importance 
to physics. Measuring its properties with precision will allow us to probe the limits of the standard model, and may point toward non-standard model physics. Using simple estimates of physics backgrounds and separable signals, we have estimated that with $4.2 \fbi$ of integrated luminosity a Muon Collider Higgs Factory, directly producing the Higgs as an $s$-channel resonance, can determine the mass of the Higgs boson to a precision of $\pm 0.06 \mev$, and its total width to  $\pm 0.18 \mev$. We estimate that, with a beam spread of $\sim 4.2 \mev$, approximately $368 {\rm~pb}^{-1}$ total integrated luminosity would be required to locate the narrow Higgs peak. We believe that these preliminary results strongly motivate further research and development towards the construction of a Muon Collider Higgs factory.

Our estimates plausibly assume there is managable machine-induced background and that the detector has excellent tracking and calorimetry. Our results underscore the value of the large, resonant Higgs cross section and narrow beam energy spread available at a muon collider. These two factors enable the direct measurement of the Higgs mass and width by scanning the Higgs s-channel resonance, which is not possible at any $e^+e^-$ collider. Our study of the physics-induced background and separation of the Higgs signal shows that significant reduction of the physics background can be achieved by a detector with high energy and spatial resolution. We believe that this justifies a more in-depth analysis of Higgs channels and their backgrounds, for example the reconstruction of $h\rightarrow WW^*\rightarrow 4j$ events using learning algorithms, or the application of flavor-tagging techniques to tag $b\bar{b}$  and $c \bar c$ events. 

While there has been considerable progress
in understanding machine-induced backgrounds, mainly from muon decays in the beam, these present a challenge which has not yet been studied in sufficient detail. We believe that, in addition to significant shielding in the detector cone and endcaps, it will be important to have a calorimeter with high spatial and temporal resolution. Our results motivate an in-depth analysis of the machine-induced background including simulation in a highly segmented, totally-active, dual readout calorimeter 
such as the MCDRCal01 detector concept.

A community letter supporting this project has been submitted previously to the archive \cite{Cline1}.

\newpage
\tableofcontents
\newpage
      

\section{Introduction}

The recent discovery of a particle with a mass of $126 \gev$ \cite{Higgs} and with a favored $J^{PC}=0^{++}$ is  widely interpreted as
a Higgs boson, as anticipated by Weinberg in the 
original incarnation of the standard model \cite{Weinberg:1967tg}.  The Higgs boson ``accommodates" the masses of quarks, leptons and electroweak gauge bosons seen in nature.  However, the origin of the Higgs-Yukawa coupling constants and mixing angles, as well as the origin of the Higgs boson mass itself, remain a mystery. The most important issues facing modern high energy physics are, therefore, to fully understand the origin of electroweak symmetry breaking and to probe for any associated new physics at the electroweak scale.

 New physics can potentially be revealed in detailed studies of Higgs boson parameters, such as the mass, decay widths and production amplitudes. Indeed, the LHC will go a long way toward revealing these detailed properties.  New physics may also be discovered at higher energy scales indirectly, \eg, through precision experiments at the ``Intensity Frontier,"  
such as through rare kaon decays, electric dipole moment searches, and 
probes of charged lepton and neutrino flavor physics. This is the purpose of ``Project-X'' in the near term at Fermilab.
However, it is also important that the field continues to evolve along the path
toward the direct probes of new physics, \ie, at the ``Energy Frontier.'' 
This demands a cost effective, and upward scalable
(in energy) strategy toward a program that can shed further light on the questions of electroweak 
physics and detailed properties of the Higgs boson.

\begin{figure}[h!] 
  \begin{center}
    \includegraphics[clip=true, scale=0.7]{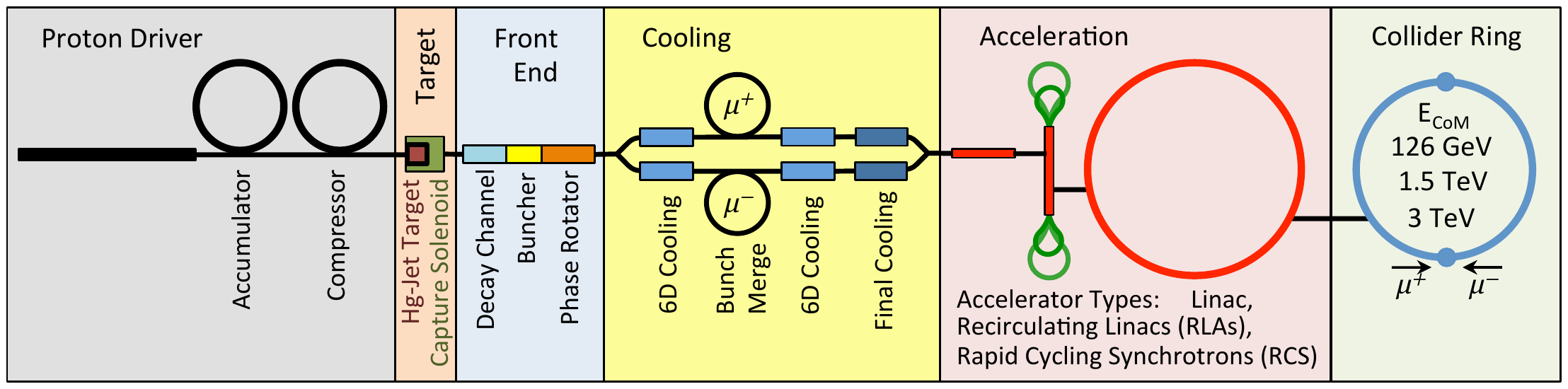}
    \caption{\small Muon Collider schematic.  A Higgs factory is simpler since it does not use final cooling and does not use a RCS.  }   
    \label{fig:lumino1} 
\end{center}    
\end{figure} 

An attractive option along this path is the development of an $s$-channel Higgs factory using muons in a compact circular collider, \ie, a ``Muon Collider Higgs Factory''\cite{Cline1, MPalmer:2013ucla}.  The muon has a Higgs-Yukawa
coupling constant that enables direct $s$-channel production of the standard model Higgs boson
at an appreciable rate. With an attainable, very small beam energy resolution
of order $\sim 4 \mev$ operating at an energy of $m_h/2 = 63$ GeV and at a nominal luminosity of about  $\sim 10^{32}$ cm$^{-2}$ s$^{-1}$
such a collider would produce $17,000$ Higgs bosons per year. 
It affords precise observation of the mass and width of the Higgs boson by
direct scanning, and the most precise determination of a Higgs-Yukawa coupling constant, that of the muon itself. 
 
\begin{table}[h!] 
 \begin{center}
   \includegraphics[clip=true, scale=0.45]{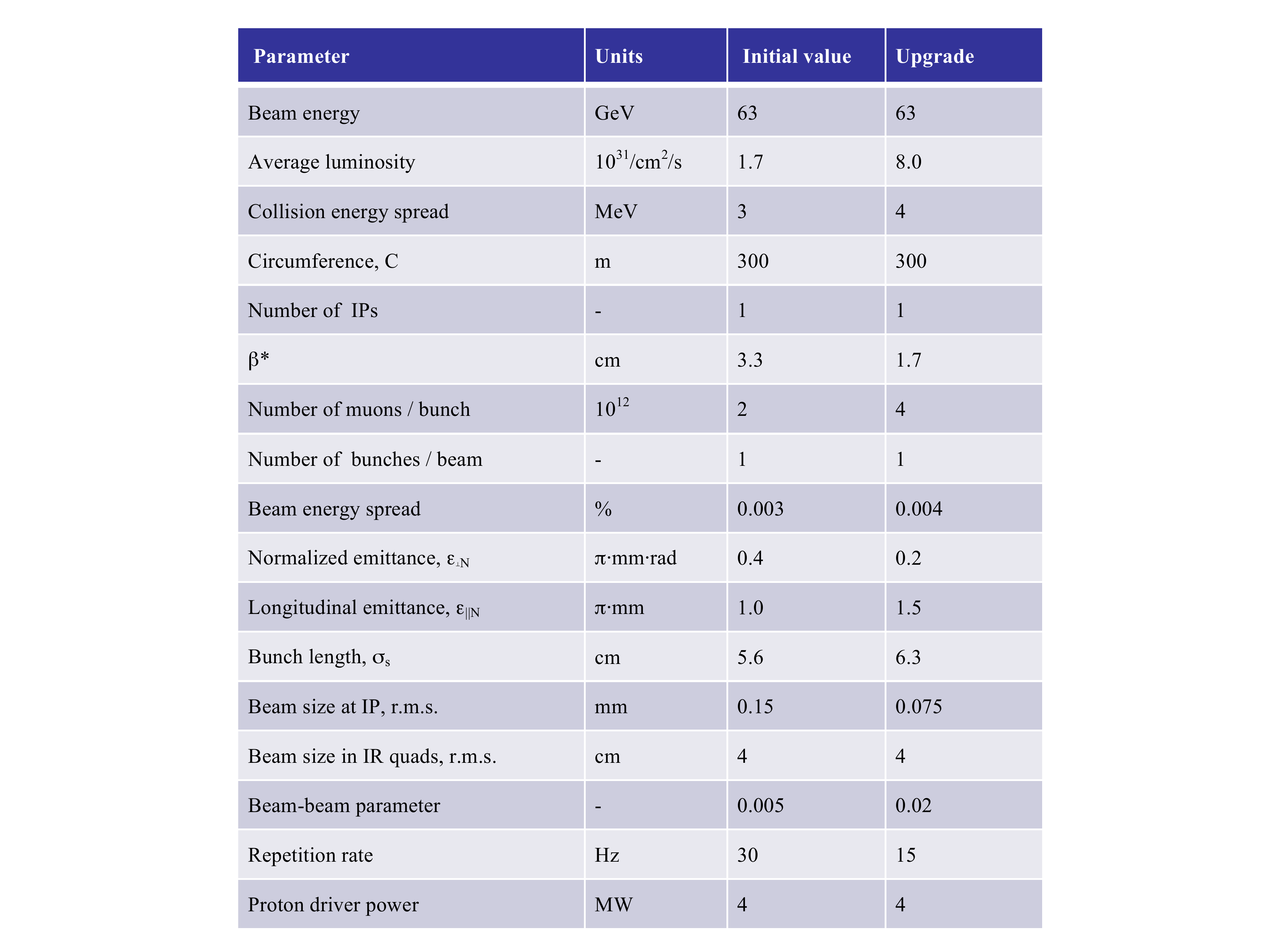}
   \caption{\small Parameters of a $\mu^+\mu^-$ Collider at 126 GeV center-of-mass energy.  The baseline design is shown.}
   \label{tab:HFparams} 
 \end{center}    
\end{table}    
     
Since the muon is about 200 times heavier than an electron, synchrotron radiation from muon beams 
in a small radius circular machine is dramatically suppressed. This allows a muon collider facility to be much smaller than an $e^+ e^-$ facility at
the same center-of-mass energy. The machine we are describing presently is 
detailed by Neuffer \cite{Neuffer:2012zze} and involves a collider storage ring of
approximately $100$ meters in diameter, roughly the size of the Fermilab Booster.  The Muon Collider also provides
superb energy resolution and an attractive way to experimentally  precisely measure the beam energy.

A conceptual design for a Muon
Collider Higgs Factory facility is shown in Fig. 1, consisting of a source of short high-intensity proton pulses, a
production target with collection of secondary $\pi$-mesons,
followed by a decay channel.
The produced $\mu^\pm$'s are collected and enter
a bunching and cooling channel. Narrow intense muon pulses are
then accelerated.

Using clever sequencing and timing, the separate $\mu^-$ and $\mu^+$
bunches can be accelerated in the Project-X linac that produces the original intense proton source, 
up to $\sim 0.5\times 10^{12}$ $\mu^+$ and $\mu^-$ bunches.  
For higher luminosities we require an extended linac of the same type used for Project X with recirculation arcs.  
The accelerated bunches are injected into the collider storage ring for collisions within an interaction region inside a detector.

\begin{figure}[h!] 
  \begin{center}
    \includegraphics[clip=true, scale=0.6]{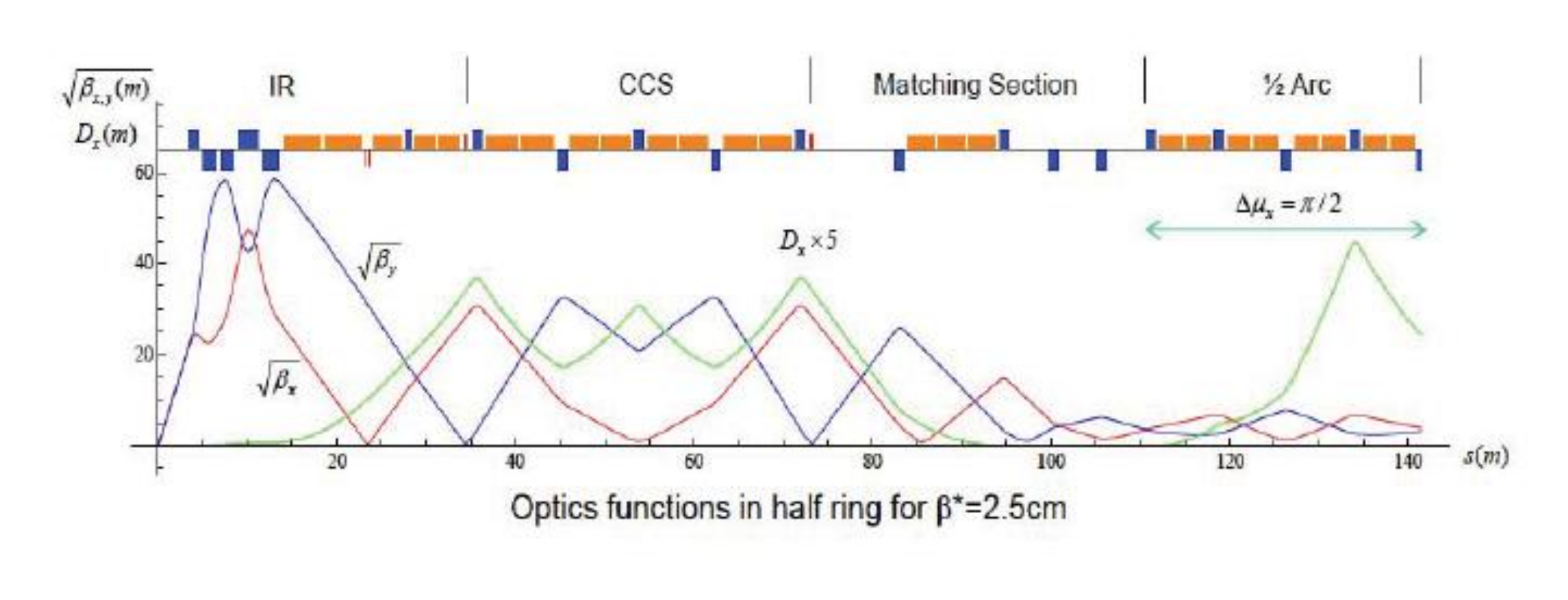}
    \caption{\small Muon Collider Higgs Factory lattice.  }   
    \label{fig:lattice} 
\end{center}    
\end{figure}

The parameters for a Muon
Collider Higgs Factory are given in Table \ref{tab:HFparams}  \cite{Alex1}.  
The lattice for the baseline design is shown in Fig. \ref{fig:lattice}. With these parameters,
luminosities of $1.7$ to $8.0$ $ \times 10^{31}$ cm$^{-2}$s$^{-1}$ can be achieved, giving 3,400 to 13,600 Higgs bosons per
year.

\newpage

\section{Physics at a Muon Collider Higgs Factory}

A Muon Collider Higgs  Factory  has the  following {\em a priori }
advantages for physics:

\begin{itemize}

\item  Small beam energy resolution  (SBER)  $\delta E/ E \lta \makebox{(few)} \times 10^{-5}$,  allowing the 
study of direct s-channel production and a line-shape scan of the Higgs boson, 
as well as other heavier Higgs bosons as in multi-Higgs models. 

\item  The s-channel Higgs production
affords  the  most  precise  measurement of a second generation
fermion Higgs-Yukawa coupling constant, 
the muon coupling, $g_\mu$, to a precision  $\delta g_\mu/g_\mu \sim \makebox{(few)} \%$.
It allows the measurement of
the renormalization group running of $g_\mu(q^2)$ from $q^2=0$  to $q^2 = m_h^2$ 
\cite{grinstein:2013ucla}.

\item The s-channel Higgs production
affords the {\em best  mass  measurement of the Higgs boson}  to a precision of $\sim \makebox{(few)} \times 10^{-6}$ with a luminosity of ${\cal{L}} \sim 10^{32} $  cm$^{-2}$s$^{-1}$.

\item It
affords  the best direct  measurement of the Higgs boson width 
to a precision of $\sim \makebox{few} $\% with a luminosity of ${\cal{L}} \sim 10^{32} $  cm$^{-2}$s$^{-1}$; see Tables [2] and [3] below. 

\item This would yield the most
precise measurement of Higgs branching ratios to $WW^*$,  $ZZ^*$  and
$b\bar{b}$.

\item At an upgraded luminosity of ${\cal{L}} \sim 10^{33} $  
cm$^{-2}$s$^{-1}$ and
$\sim 3$ ``snowmass years"  on the Z-pole,  we would
produce  $\sim 10^9$, Z-bosons; thus  
the Muon Collider Higgs Factory upgrade permits a ``Giga-Z  program." 

\end{itemize}

Detailed studies of these and other issues  are underway,
including: (1)  
optimal search strategy for the Higgs peak establishing  
a threshold integrated luminosity for
physics of about $\sim \makebox{(few)} \times 10^{31}$ cm$^{-2}$s$^{-1}$
\cite{Conway:2013lca, Eichten:2013ucla} ; (2)
charm  decay mode  $h\rightarrow c\bar{c}$ appears to be accessible  
at a level of $\sim 8\sigma$ \cite{Purohit:2013ucla} ; 
(3) possible observable interference effects may occur in
\eg,  $h\rightarrow WW^*$;
(4) the mode $h\rightarrow Z \gamma^*\rightarrow Z \ell\ell$ may prove to be an 
interesting mode for study the decay spectrum.

In Section (5) we will discuss the physics and detector issues for many
standard Higgs boson processes. We turn presently to a brief discussion of
the location and line-shape scan of the Higgs boson, produced directly in the $s$-channel, at a Muon Collider Higgs Factory.


\subsection{Higgs Boson Signal and Background}
\label{sec:sigbkg}

One of the most appealing features of a muon collider Higgs factory is its $s$-channel resonant production of Higgs bosons. For the production $\mm \to h$ and a subsequent decay to a final state $X$ with a $\mm$ (partonic) c.m.~energy
$\sqrt{\hat s}$,  the Breit-Wigner resonance formula is:
\bea
\sigma(\mm \to h \to X)
= \frac{4\pi \Gamma_h^{2} \br(h\to \mm )\br(h\to X )} {(\hat s-m^2_h)^2 + \Gamma_h^2 m_h^2},
\label{eq:sigma}
\eea
At a given energy, the cross section is governed by three parameters: $m_h$ for the signal peak position, $\Gamma_h$ for the line shape profile, and the product $B \equiv {\br}(h\to \mu^+\mu^- ){\br}(h\to X )$ for the overall event rate.

In reality, the observable cross section is given by the convolution of the energy distribution delivered by the collider with eq(\ref{eq:sigma}). Assuming that the $\mm$ collider c.m.~energy ($\sqrt s$) has a luminosity distribution
$$ { dL(\sqrt s) \over d\sqrt{\hat s} }
= {\frac 1 {\sqrt{2\pi \Delta}} } \exp[\frac {-( \sqrt{\hat s} - \sqrt s)^2} {2\Delta^2}] ,
$$
with a Gaussian energy spread $\Delta  = R \sqrt {s}/\sqrt{2}$, where $R$ is the relative beam energy resolution, then the effective cross section is
\bea
&& \sigma_{\rm eff}(s) = \int d \sqrt{\hat s}\  \frac {dL(\sqrt s)} {d\sqrt{\hat s}}   \sigma(\mm \to h \to X)
\label{eq:convol}
\eea

The excellent beam energy resolution of a muon collider allows a direct determination of the Higgs boson width, in contrast to the situations 
in the LHC and ILC~\cite{Neuffer:2012zze}. 
We calculate the effective cross sections at the peak for two different energy resolutions, $R=0.01\%$ and $R=0.003\%$.
We further evaluate the signal and SM background for the leading channels, $h \to b \bar b,~WW^*$.

We impose a polar angle acceptance for the final-state particles, $10^\circ < \theta  < 170^\circ$.
We assume a $60\%$ single $b$-tagging efficiency and require at least one tagged $b$ jet for the $b\bar b$ final state.
The backgrounds are assumed to be flat with cross sections evaluated at $126~\gev$ using Madgraph5 \cite{Alwall:2011uj}.
\begin{table}[tb]
\centering
\begin{tabular}{|c|c|c|c|c|c|}
  \hline
 & $\mm\rightarrow h$ & \multicolumn{2}{|c|}{$h\rightarrow b\bar b$} &  \multicolumn{2}{|c|}{$h\rightarrow WW^*$}  \\ \cline{3-6}
 \raisebox{1.6ex}[0pt]{R (\%)} & $\sigma_{\rm eff}$ (pb) & $\sigma_{Sig}$ & $\sigma_{Bkg}$ &  $\sigma_{Sig}$ & $\sigma_{Bkg}$ \\ \hline
 $0.01$ & $16$ & $7.6$ & & $3.7$ &  \\ \cline{1-3} \cline{5-5}
$0.003$ & $38$ & $18$ & \raisebox{1.6ex}[0pt]{$15$}  & $5.5$ & \raisebox{1.6ex}[0pt]{$0.051$}\\ \hline
\end{tabular}
\caption{Effective cross sections (in pb) at the resonance $\sqrt {s}=\mh$ for two choices of beam energy resolutions $R$ and two leading decay channels, with the SM Higgs branching fractions $\br_{b\bar b}=56\%$ and $\br_{WW^*}=23\%$~\cite{Dittmaier:2012vm}. This table is taken from Ref~\cite{Han:2012rb}.}
\label{tab:cpara}
\end{table}

Results are tabulated in Table \ref{tab:cpara}.
The background rate of $\mm\rightarrow Z^*/\gamma^* \to b\bar b$ is $15~\pb$, and the rate of $\mm \to WW^*\to 4$ fermions is only $51~\fb$, as shown in Table~\ref{tab:cpara}. Here we consider all decay modes of $WW^*$ because of its clear signature at a muon collider.
The four-fermion backgrounds from $Z\gamma^*$ and $\gamma^*\gamma^*$ are initially smaller, but can be greatly reduced by kinematical 
constraints, such as requiring the invariant mass of one pair 
of jets to be near $m_W$ and setting a lower cut 
on the invariant mass of the other pair.

\subsection{Locating the Mass Window of the Higgs Boson}
\label{sec:window}

The first challenge at the muon collider
is to locate the Higgs boson mass within a fine mass window.
The theoretical
natural width of a $125 \gev$  Higgs is $4.07 \mev $.
 It is expected that the Higgs mass will be known 
 to better than $\sim 100 \mev $ from measurements at the LHC.  
 After the Higgs has been located, the Muon Collider
 can be used to scan the line-shape and then tuned to sit on the peak of the Higgs resonance to study Higgs physics.
 
 A. Conway, H. Wenzel \cite{Conway:2013lca} and E. Eichten \cite{Eichten:2013ucla} have developed the optimal 
 strategy to locate the Higgs boson at a Muon Collider.  
The required total luminosity is determined to observe a $5\sigma$ ($3\sigma$) Higgs boson signal with various likelihoods and energy steps.  It was assumed that a center-of-mass energy resolution of $3.54$ MeV is obtainable at a Muon Collider.  
 
 Two decay channels were considered: (1) the $b \bar b$ final state and  (2) the $W W^*$  final state.   
The $b\bar b$ final state has the largest branching fraction, but has a significant background, even with the excellent 
 energy spread possible at the muon collider.  The $WW^*$ channel has very small background physics rates (two orders of magnitude smaller than the Higgs signal). For this study the $WW^*$ decays with one neutrino-lepton pair are used.

The best strategy was found to use energy steps equal to the beam energy spread with the choice of the next energy bin  determined  by
the maximum probability of discovery weighted by the prior of the LHC measurements.  The $WW^*$ channel is the most effective  but
both channels are included in the final results.
The results for the combined channels from Conway and Wenzel \cite{Conway:2013lca} is shown in  Table \ref{tab:findH} for various p-values of non-discovery. 

\begin{table}
     \begin{centering}
	\begin{tabular}{|c|l|r|r|}
		\hline
		\multirow{2}{*}{Channel}	& $\sigma_{sig}\ (pb)$	& \multicolumn{2}{|c|}{Luminosity Required $(pb^{-1})$} \\ \cline{3-4}
		& $\sigma_{bkgr}\ (pb)$	& $ 3\sigma$ signal & $ 5\sigma$ signal\\ \hline
		\multirow{2}{*}{$b\bar{b}$ (Cut)} & $\sigma_s = 4.0$ & 
			\multirow{2}{*}{1,100}	& \multirow{2}{*}{2,200} \\
			& $\sigma_b= 6.3$	&	& \\ \hline
		\multirow{2}{*}{$WW^*$ (Cut)} & $\sigma_s = 1.9$ & 
			\multirow{2}{*}{400}	& \multirow{2}{*}{1,100} \\
			& $\sigma_b= 0.001$	&	& \\ \hline
		\multirow{2}{*}{$b\bar{b},\ WW^*$} & $\sigma_s = --$ & 
			\multirow{2}{*}{600}	& \multirow{2}{*}{900} \\
			& $\sigma_b= --$	&	& \\ \hline
	\end{tabular}
	 \caption{Required luminosity to guarantee finding a $3$ or $5\sigma$ Higgs signal with confidence level $p = 3\sigma$ 
	    \cite{Conway:2013lca}.}
	 \label{tab:findH}
   \end{centering}
\end{table}

It is clear from Table \ref{tab:findH} that reducing the error on the LHC determination of the mass of the Higgs could greatly aid 
finding the Higgs resonance at a Muon Collider.

\subsection{Precision Determination of the Total Width}
\label{sec:width}

The total decay width ($\gmh$) is one of the most important of all of the Higgs boson properties.  Once the total width is known, the measurement of the partial decay widths to different channels yields model-independent determination of Higgs boson coupling strengths.

\begin{figure}[htb]
\begin{center}
\subfigure{
      \includegraphics[width=130pt]{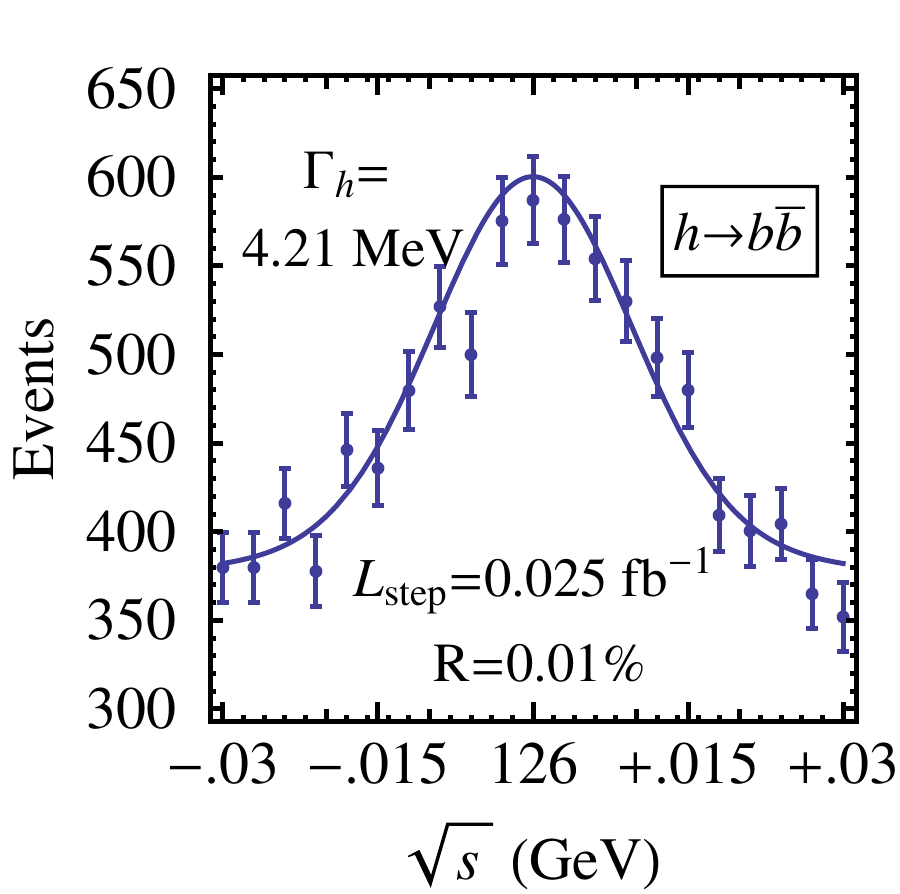}}
\subfigure{
      \includegraphics[width=130pt]{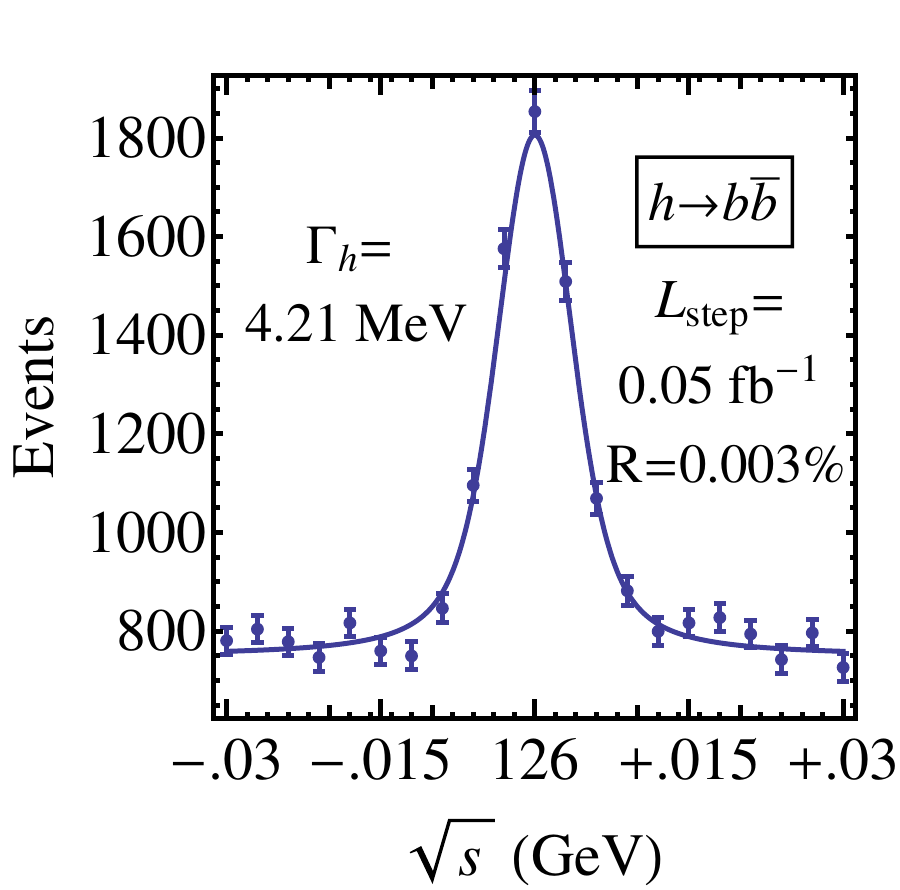}}
\subfigure{
      \includegraphics[width=130pt]{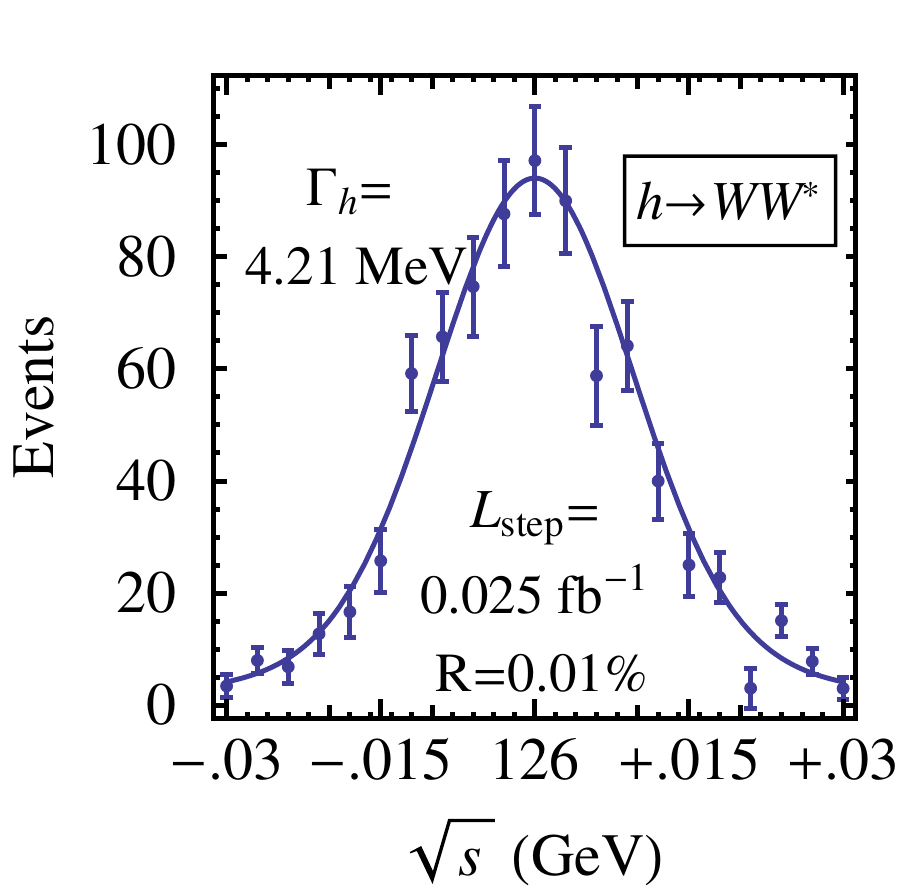}}
\subfigure{
      \includegraphics[width=130pt]{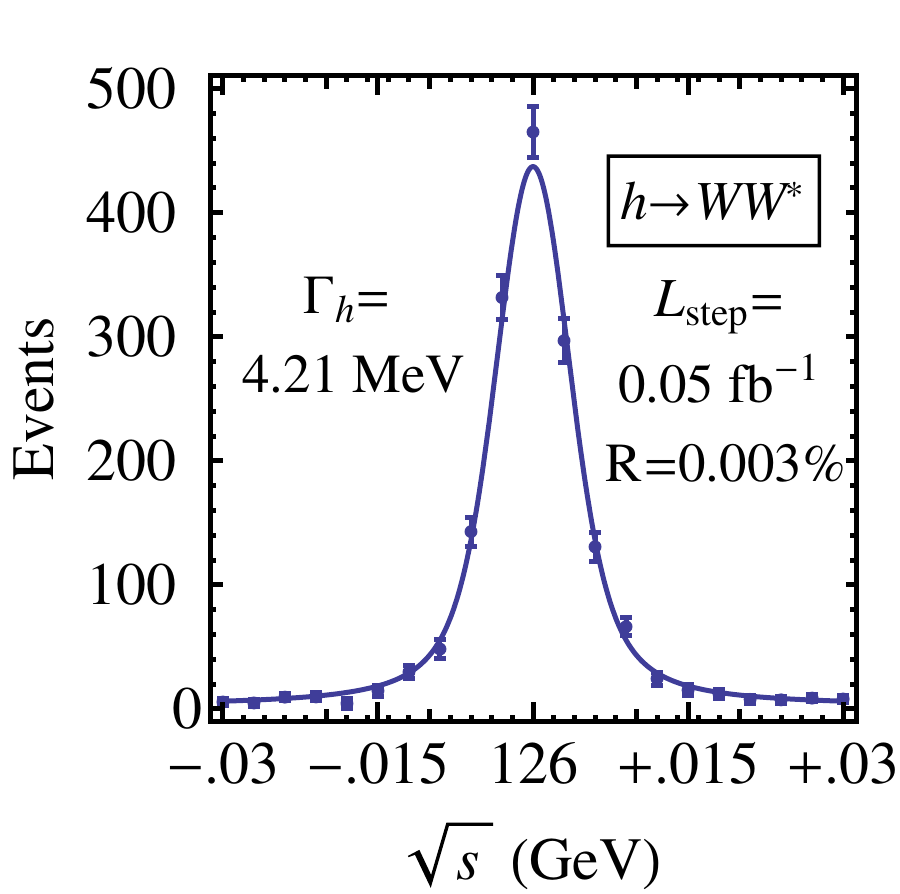}}
\caption{Number of events of the Higgs signal plus backgrounds and statistical errors expected for two different beam energy resolutions and integrated luminosities as a function of the collider energy $\sqrt s$ in $b\bar b$ and $WW^*$ final states with a SM Higgs $\mh =126~\gev$ and $\gmh = 4.21~\mev$. Detector backgrounds are not included, see more discussion in Sec. \ref{sec:backgrounds}. These figures are taken from Ref~\cite{Han:2012rb}.}
\label{fig:shape12}
\end{center}
\end{figure}

We first generate pseudo-data in accordance with a Breit-Wigner resonance at $126~\gev$ convoluted with the beam energy profile integrated over $\sqrt {\hat s}$.
These data are then randomized with a Gaussian fluctuation with variance equal to the total number of events expected, including both signal and background. The results for the leading two channels $b\bar b$ and $WW^*$ are shown in Fig.~\ref{fig:shape12} for different integrated luminosities.

\begin{figure}[htb]
\begin{center}
\subfigure{
      \includegraphics[width=145pt]{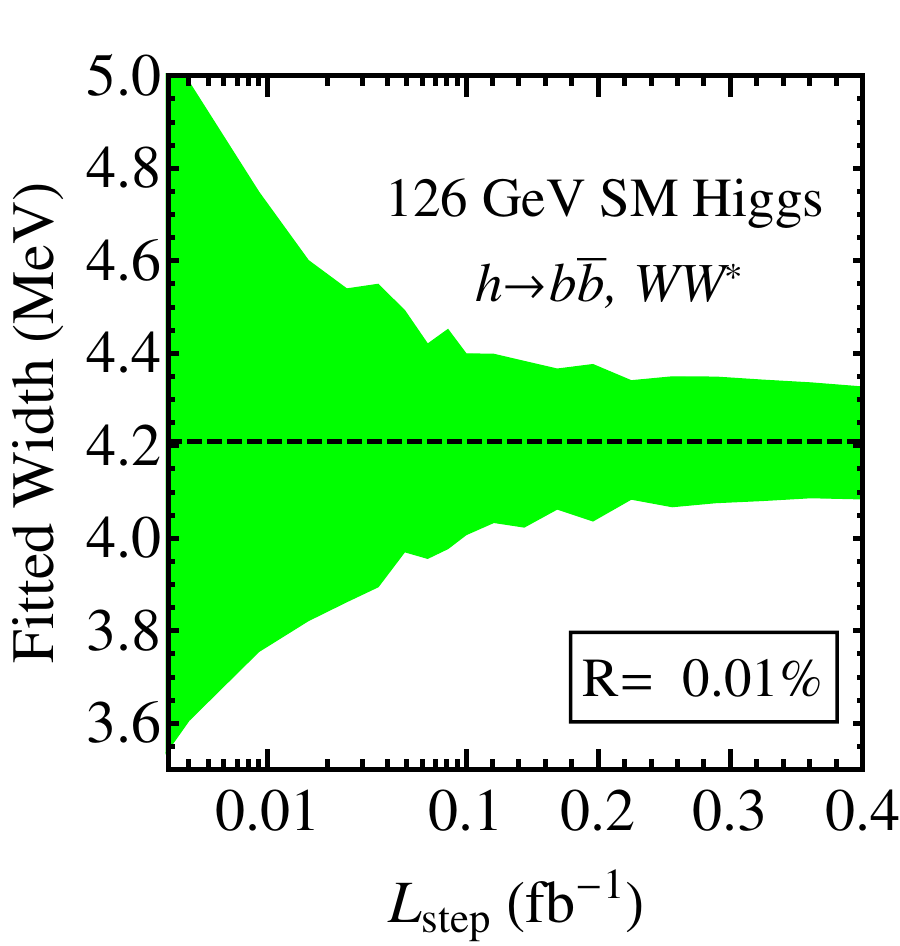}}
\subfigure{
      \includegraphics[width=145pt]{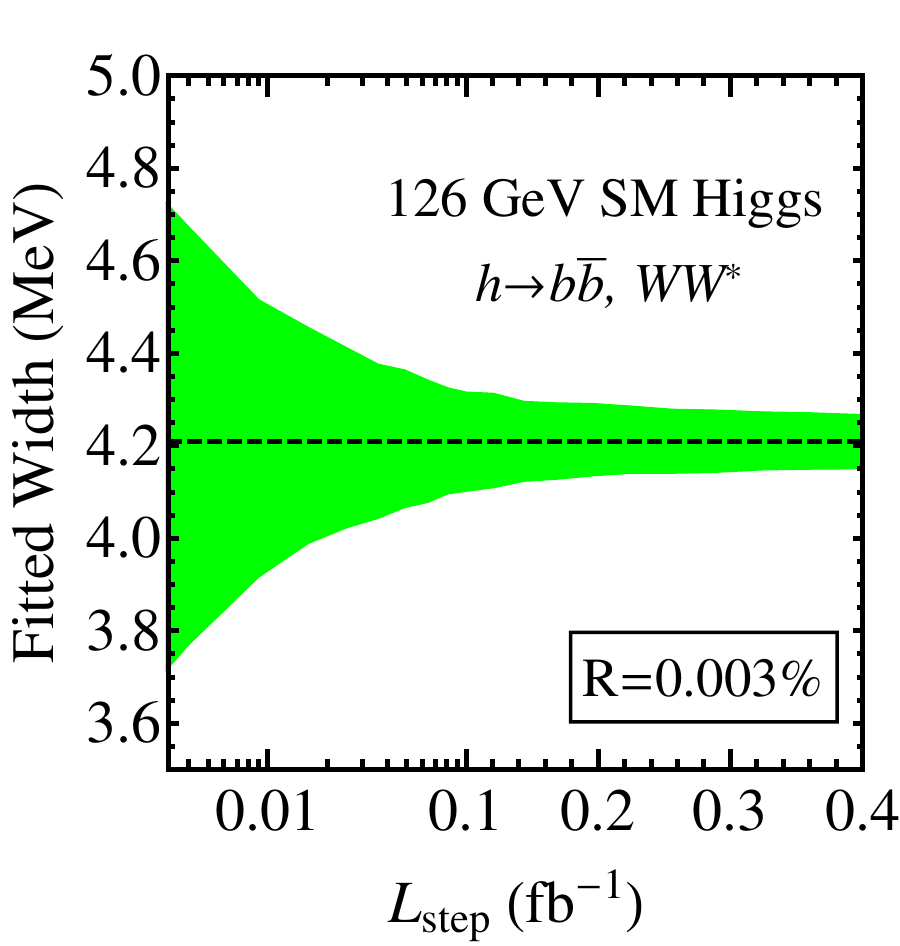}}
\caption{Fitted values and errors for the SM Higgs width versus the luminosity per step with the 20-step scanning scheme with 3-MeV step size. These figures are taken from Ref~\cite{Han:2012rb}.
}
\label{fig:luminofit}
\end{center}
\end{figure}

We adopt a $\chi^{2}$ fit over the scanning points 
with three model-independent free parameters in the theory
$\gmh,\ B$ and $\mh$ as in Eq.~\ref{eq:sigma}.
We show our results for the SM Higgs width determination in
Fig.~\ref{fig:luminofit} for beam resolutions, $R=0.01\%$ and $R=0.003\%$  by varying the luminosity.
The achievable accuracies with the 20-step scanning scheme by combining two leading channels are summarized in Table~\ref{tab:acrcy} 
for three representative luminosities per step.

\begin{table}[htb]
\centering
\begin{tabular}{|c|c|c|c|c|}
  \hline
  $\Gamma_h=4.21~\mev$ & $L_{\rm step}$ ($\fbi$) & $\delta{\Gamma_h}~(\mev)$ & $\delta B$ & $\delta m_{h}~(\mev)$ \\ \hline \hline
  & $0.005$ & $0.73$ & $6.5\%$  & $0.25$ \\ \cline{2-5}
  $R=0.01\%$ & {\bf 0.025} & {\bf 0.35} & {\bf 3.0\%} & {\bf 0.12} \\  \cline{2-5}
  & $0.2$ & $0.17$ & $1.1\%$ & $0.06$ \\ \hline \hline
  & $0.01$ & $0.30$ & $4.4\%$ & $0.12$ \\ \cline{2-5}
  $R=0.003\%$ & {\bf 0.05} & {\bf 0.15} & {\bf 2.0\%} & {\bf 0.06} \\ \cline{2-5}
  & $0.2$ & $0.08$ & $1.0\%$ & $0.03$ \\
  \hline
\end{tabular}
\caption{Fitting accuracies for $\Gamma_h$, $B$, and $m_h$ of the SM Higgs with the 20-step scanning scheme with $3~\mev$ step size for three representative luminosities per step. Results with total integrated luminosity $0.5~\fbi$ ($1~\fbi$) for resolution $R=0.01\%(0.003\%)$ are in boldface. This table is taken from Ref~\cite{Han:2012rb}.}
\label{tab:acrcy}
\end{table}

The mass and cross section can be simultaneously determined along with the Higgs width to a high precision.
The results obtained are largely free from theoretical uncertainties.
 The major systematic uncertainty comes from our knowledge of beam properties \cite{Neuffer:2012zze}. The uncertainty associated with the beam energy resolution $R$ will directly add to our statistical uncertainties of Higgs width. This uncertainty can be calibrated by experimentalists. On the other hand, the beam profile is unlikely to be Breit-Wigner resonance profile. Thus an additional fitting parameter of the beam energy distribution is anticipated to provide us additional knowledge about the beam energy. Our estimated accuracies are by and large free from detector resolutions. Other uncertainties associated with b tagging, acceptance, etc., will enter into our estimation of signal strength $B$ directly. These uncertainties will affect our estimation of total width $\Gamma_h$ indirectly through statistics, leaving a minimal impact in most cases.

\newpage

\newpage

\section{Collider Environment}


\subsection{Machine Performance and Environment}

A preliminary machine design for a Muon Collider Higgs Factory was
presented at the
Muon Accelerator Program collaboration meeting in June,
2013~\cite{alexahin}.
This is a unique machine, designed to provide high luminosity, very
small energy
spread, excellent stability, and good shielding of muon beam decay
backgrounds.

There are several features of the accelerator which impact the detector.
The ring is 300 meters in circumference, corresponding to a 1 $\mu$s
crossing interval.
Beams are injected at 30 Hz and circulate for roughly 1000 turns per store.
The lattice is designed both to provide high average luminosity
($1.7\times10^{31}$ cm$^{-2}$  s$^{-1}$)
and very small ($3\times10^{-5}$) energy spread. The luminosity is
achieved utilizing
a quadruplet final focus
with very large aperture (50 cm) quadruple magnets at the interaction region.
The large beam radius, 5.6 cm bunch length,  and proximity (3.5 meters)
of the last quadrupole to
the interaction region will require a redesign of  the beam background
shielding with respect to the existing
$1.5 \tev$ machine design. A final design will likely involve a compromise
between
luminosity, detector acceptance, and machine backgrounds. In this note
we utilize the
backgrounds that have been generated and studied for the $1.5 \tev$ machine.

\subsection{Energy Determination by Spin Tracking}

In the present scenario
the $\mu $ beams are created with a small polarization ($\approx 10$ to $20\% $)
from a bias toward capture of forward $\pi \rightarrow \mu $ decays. 
This polarization should be substantially maintained through the
cooling and acceleration systems leading to the circular storage ring of the muon collider.

While circulating in a Higgs factory storage ring, as described in
the previous sections,  muons continuously decay 
at a rate of $\cong   10^7$ decays per meter.  As a result of 
the muon polarization, electrons and positrons 
from the muon decays will have a mean energy that is dependent upon
the polarization of the muons.  That polarization, $\bf{P}$
will precess as the beam rotates around the ring. The precession
will, in turn, modulate the mean energy of decay electrons, and therefore 
provide a modulated signal at a detector capturing those decays.  

The mean energy from decay electrons is:
\begin{equation}
<E(t)>\; = \;<(\frac{7}{20}N E_\mu \exp({-\alpha t}))(1+\frac{\beta}{7}\bf{P}\cos(\omega t+\phi) )>
\end{equation}
where $N$ is the
initial number of $\mu $'s,
$E_\mu$ is the $\mu $ energy, ${\alpha}$ is the decay parameter,
$\beta = v/c$, $\bf{P}$ is polarization, $\phi $ is a phase, and $t$ 
is time in turn numbers, and:
\begin{equation}
\omega=2\pi\lambda(\frac{g-2}{2})\approx 2\pi\times0.7
\end{equation}
is the precession frequency that depends upon the muon beam energy.  
A detector capturing a significant number of decay electrons will have a
signal modulated by $\omega$. The frequency can be measured to
very high accuracy, leading to a muon beam energy measurement at the
$10^{-6}$ level of precision (corresponding to $\sim 0.1 \mev$), or
better. 

The precession observation determines the muon energy at each individual
collision store. The precession signal decreases with time following
the muon decay and the energy width, providing an important measurement
of that width, which will assist in unfolding the Higgs width.

Raja and Tollestrup~\cite{pol1, pol2}, and Blondel~\cite{pol3}, have analysed the polarization-based energy determination for  a muon
collider. Blondel established that a polarization of 5\%
would be sufficient to enable this measurement~\cite{pol4}. Simulations indicate
that the muons should have an initial polarization of $(10 -20) \%$
and that polarization would be maintained in cooling and acceleration.


\subsection{Accelerator Backgrounds \label{sec:backgrounds}}

The ability to perform physics measurements
with a muon collider will largely be
determined by how well one can suppress the accelerator backgrounds
from the collider ring.  The source of most of
the accelerator backgrounds in a muon collider is associated with the
decay of beam muons.  Studies to date have utilized backgrounds generated 
for a $1.5 \tev$ collider \cite{M1}.  Specific backgrounds for a $126 \gev$ Higgs factory  have been recently calculated, \cite{M2, M3}.
Both the $1.5 \tev$ machine and Higgs factory designs assume
$2 \times 10^{12}$ muons per bunch, which will produce
$4.3\times 10^{5}$ or $6\times 10^{6}$ muon decays per meter for 
the $1.5 \tev$ and $126 \gev$ machines. 

The electrons resulting from muon decays will interact with
the walls of the beam chamber, collimators, and shielding, 
 producing high energy electromagnetic
showers, synchrotron radiation, photonuclear interactions, and 
Bethe-Heitler muons. Photonuclear interactions with the nuclei of beam pipe, magnet or
shielding material from energetic photons in the electromagnetic
shower are the main source of the hadronic and neutron background.
Neutrons are predominantly produced from photonuclear spallation
processes in the giant resonance region ($14-20 \mev$ incident
gammas).

A preliminary study of backgrounds in a $1.5 \tev$ ($750 \gev$ on $750 \gev$) muon
collider was done utilizing MARS15, \cite{M4} \cite{M5}, and G4Beamline \cite{M6} codes. The goal of the study was to calculate the accelerator-generated
backgrounds that could arrive at a muon collider detector. The lattice design for this
machine is described in Ref. \cite{M7}, with a description of the interaction
region design given in Ref.  \cite{M8}.  In this study the lattice was
modelled at $\pm$75 m from the interaction point.  Electrons from muon decays are assumed to 
originate at locations uniformly distributed along the $\mu^+$ and $\mu^-$
reference trajectories.

\begin{figure*}[ht]
\centering
\includegraphics[width=135mm]{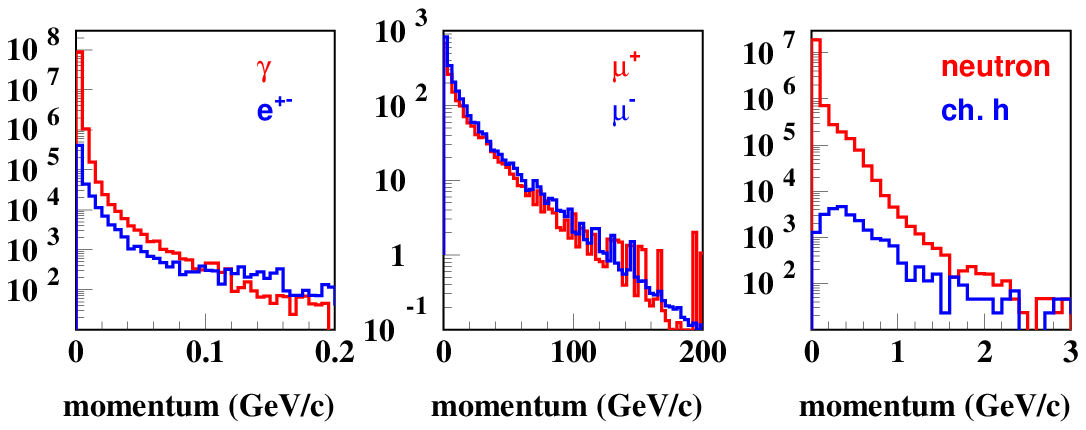}
\caption{Energy distributions of particles entering the detector region from a MARS simulation 
of $1.5 \tev$  Muon Collider beam backgrounds \cite{M1}.}
\label{fig:background1}
\end{figure*}

Figure \ref{fig:background1}  shows the background flux entering the detector region  in a typical 
Muon Collider interaction \cite{M1}.  Total non-ionizing background is about 10\% that of the LHC, 
but the crossing interval is 400 times longer, resulting in high instantaneous flux.  
The background is very different in character than that of either the LHC or CLIC.
It is dominated by soft photons and low energy neutrons emerging from the shielding surrounding the 
detector.  A typical background event has $164 \tev$  of photons, 
$172 \tev$  of neutrons, and $184 \tev$ of muons. With the exception of muons and charged hadrons the 
background spectrum is dominated by low energy particles.
Only a small fraction of the background originates from the vicinity of the interaction region.  
This means that most of the decay background 
is out of time with respect to particles originating from the $\mu^+\mu^-$ collision.

\newpage



\section{Detector Design}
\label{sec:detector}

The next generation of collider detectors will emphasize precision for all sub-detector
systems. In the case of a muon collider Higgs factory 
the $h \rightarrow  b \bar{b}$ and $h \rightarrow  \tau^+ \tau^-$ channels require good b-tagging and vertexing capabilities.  
The $ h \rightarrow WW^*$  channel will require the capability 
to distinguish $W$ and $Z$ vector bosons in their hadronic decay mode while $h \rightarrow  \gamma \gamma$ emphasizes 
excellent energy and position measurement of photons\footnote{The discovery of the Higgs 
boson in the $h \rightarrow  \gamma \gamma$ channel at 
the LHC provides a strong argument requiring  good energy resolution
for photons.} .

\begin{figure}[htb] 
  \begin{center}
    \includegraphics[clip=true, scale=0.3]{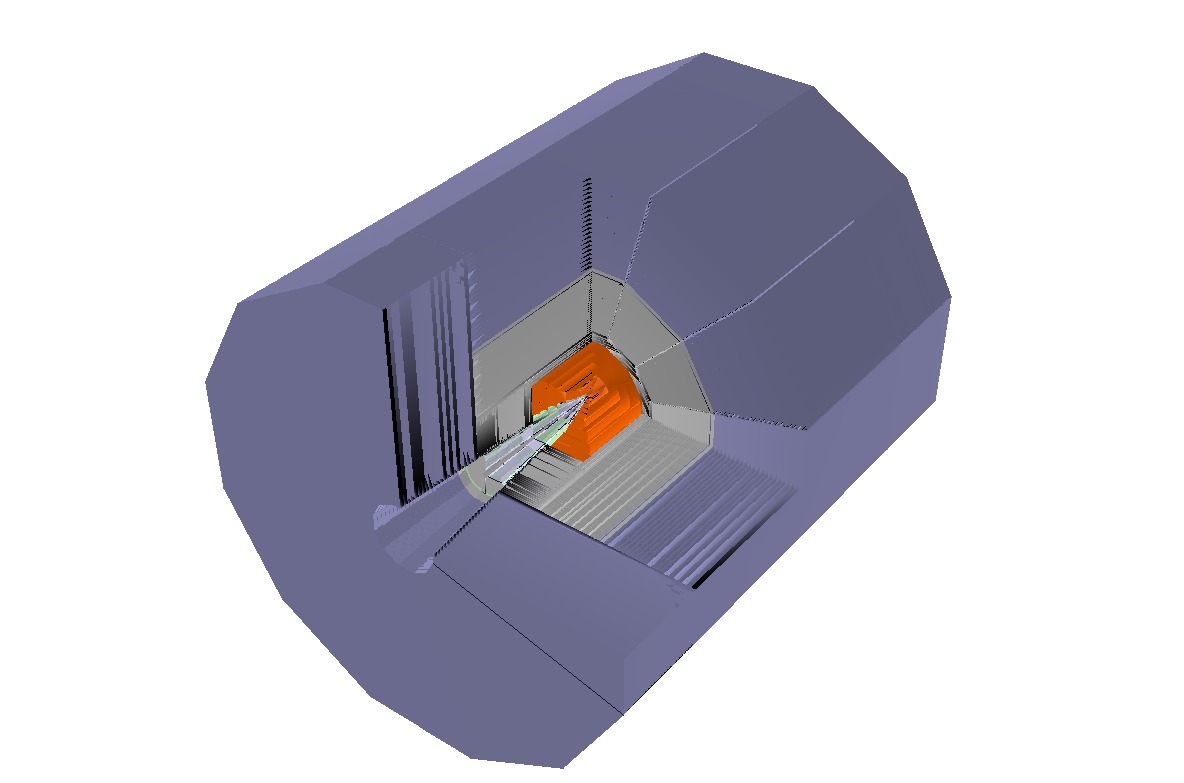}
    \caption{\small  Illustration of the mcdrcal01 detector.}   
    \label{fig:detector} 
\end{center}    
\end{figure}

To achieve the tracking goals we require a high solenoidal magnetic
field of 5 Tesla and high precision low mass tracking. To achieve good jet resolution 
the electromagnetic and hadronic calorimeter have to be located within the solenoid.
 The machine induced background from $\mu$ decays upstream and downstream 
 of the interaction point provides a very challenging
 environment; but this background is ``out of time'' compared to the
 particles from the interaction point. 
Therefore, for both tracker and calorimeter, good timing resolution (of
order of nanoseconds) will be crucial to reduce this background to an
acceptable level. This background makes shielding necessary  extending into the tracker
volume.   Figure \ref{fig:detector} shows an illustration of the detector as it
 is currently implemented in the Geant4 simulation. The tungsten
shielding cone is quite visible. Here we present an idealistic conceptual design 
that will have to be replaced by an optimized, more realistic and cost
efficient design in the future. 

\subsection{Tracking}

\begin{figure}[htb] 
  \begin{center}
    \includegraphics[clip=true, scale=0.4]{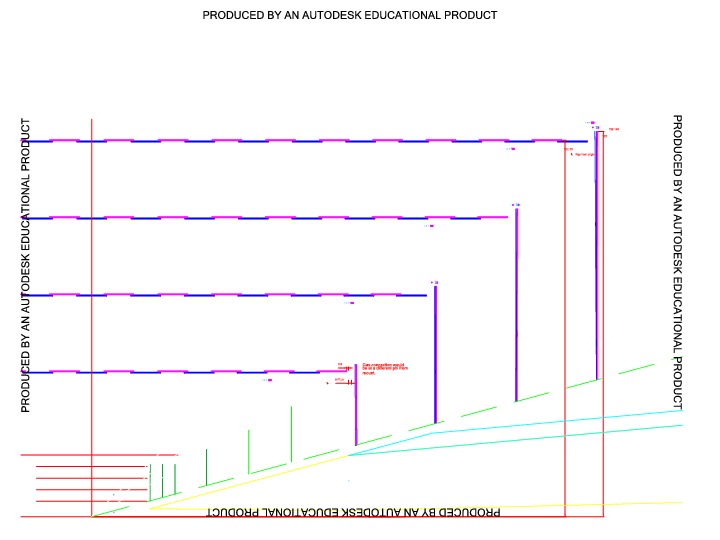}
    \caption{\small  Layout of the vertex and tracking detector barrel and endcaps and the tungsten shielding cone.}   
    \label{fig:layout} 
\end{center}    
\end{figure}

To achieve the tracking goals while keeping the tracker compact we require a high solenoidal magnetic
field of 5 Tesla and use silicon-based tracking with a pixelated vertex
detector for  high precision, low mass tracking. Fast timing and fast readout require extra power and cooling, and R\&D will be necessary to
 achieve this while keeping detectors and support at the required low mass. 
Figure \ref{fig:layout} shows the layout of the tracking and vertex detector. The vertex barrel detector is assumed to consist of 
5 barrel layers with $20$ $\mu$m square pixels and $0.8\%$ radiation length per layer,
the six vertex disks are assumed to utilize $50$ $\mu$m square pixels with $1.0\%$ radiation length. 
The four tracker barrel layers and four disk layers are assumed to have 
 $100\times 1000$  $\mu$m   short strips with $1.5\%$ radiation length per layer. 
  The large background of non-ionizing radiation means that the 
silicon tracker will have to be kept cold, around $-20$ $^o$C, to avoid radiation damage, increasing the detector mass.  
Precise timing and pixelated detectors will be crucial to a successful Muon Collider detector.  
Both come at a cost.  Fast electronics will necessarily dissipate significant power and,
in contrast to planned ILC detectors, detectors for the Muon Collider will have to be liquid (CO$_2$) cooled 
with an associated increase in mass with respect to ILC trackers.  

\subsection{Calorimetry}

A common benchmark for ILC detectors is to distinguish W and Z vector bosons in
their hadronic decay mode. This requires a di-jet mass resolution better than the natural width of
these bosons and hence a jet energy resolution better than 3\%. For hadron calorimetry this
implies an energy resolution a factor of at least two better than previously achieved to date by
any large-scale experiment. 

A novel approach to achieving superior hadronic energy
resolution is based on a homogeneous hadronic calorimetry (HHCAL) detector concept,
including both electromagnetic and hadronic parts. This employs 
separate readout of the Cerenkov and
scintillation light and using their correlation 
to obtain superior hadronic energy resolution \cite{[2],[3]}. This HHCAL detector concept has a total absorption nature, so its energy resolution is not
limited by the sampling fluctuations. It has no structural boundary between the ECAL and
HCAL, so it does not suffer from the effects of dead material in the middle of hadronic
showers. In addition there is no difference in response since the ECAL and HCAL are identical and only the 
segmentation differs. It also takes advantage of the dual readout approach by measuring both Cerenkov
and scintillation light to correct for the fluctuations caused by the nuclear binding energy
loss, so a good energy resolution for the hadronic jets can be achieved \cite{[4],[2],[3]}.

To improve event reconstruction we plan to use particle flow algorithms and therefore we require fine 
segmentation (granularity) of the calorimeter. 
A cost effective active material is crucial for the HHCAL detector concept and R\&D is necessary to find the appropriate active materials, such as
 scintillating crystals, glasses or ceramics to be used to construct an HHCAL. With regards to photosensors,  silicon-based photo detectors (a.k.a SiPM, MPPC) 
are reaching a very mature state and are becoming potential photo-transducers for hadron calorimetry for selectively detecting scintillation and Cerenkov light. 
The parameters and segmentation of the mcdrcal01 calorimeter are listed in 
Table \ref{tab:Cal-Segmentation}.
\begin{table*}[h]
 \begin{center}
 \begin{tabular}{|l|c|c|c|}
  \hline 
          & electromagnetic (em)  & hadronic (had)  & muon \\
  \hline
  \hline
  Material                               & \multicolumn{2}{c|}{BGO/PbF$_2$/PbWO$_4$}   &   Iron     \\
\hline 
  density [g$/$cm$^3$]                     &  \multicolumn{2}{c|}{$7.13/7.77/8.29$}  & $7.85$  \\
\hline
  radiation length [cm]                &  \multicolumn{2}{c|}{$1.1/0.93/0.89$} & $1.76$  \\
\hline
  nuclear interaction length (IA) [cm] &  \multicolumn{2}{c|}{$22.7/22.4/20.7$} & $16.8$  \\
\hline
  Number of layers                       & $10 $          & $30$          & $22$    \\
\hline  
Thickness of layers [cm]             & $2 $           & $5$           & $10$    \\
\hline
  Segmentation [cm $\times $ cm]          & $1\times 1$  & $2\times 2$ & $10 \times 10$ \\
\hline
  total depths [cm]                    & $20 $          & $150 $        & $220 $  \\
\hline
  total IA                               &  \multicolumn{2}{c|}{em + had: $ 7.5/7.6/8.2$} & $13.1$ \\                       
  \hline
  \end{tabular}
 \end{center}
\caption{Properties of calorimeter and instrumented Iron flux return for barrel and endcaps
for some crystal materials under consideration.}
\label{tab:Cal-Segmentation}
\end{table*}



\begin{figure}[htb] 
  \begin{center}
    \includegraphics[clip=true, scale=0.3]{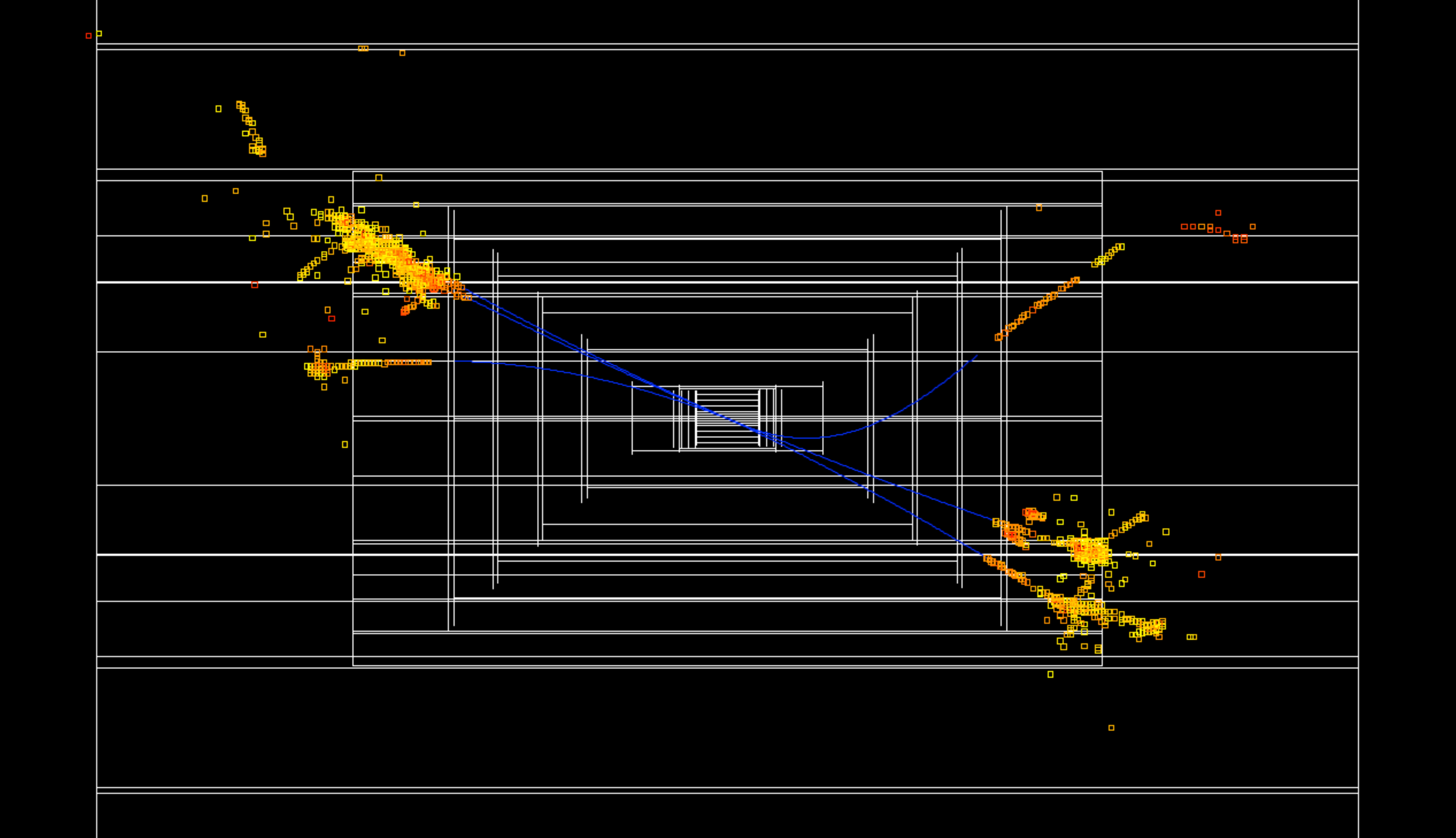}
    \caption{\small  Wired display of a $h\rightarrow \tau^+\tau^-$ event in the mcdrcal01 detector.}   
    \label{fig:h-tautau-1l} 
\end{center}    
\end{figure} 

\begin{figure}[htb] 
  \begin{center}
     \includegraphics[clip=true, scale=0.3]{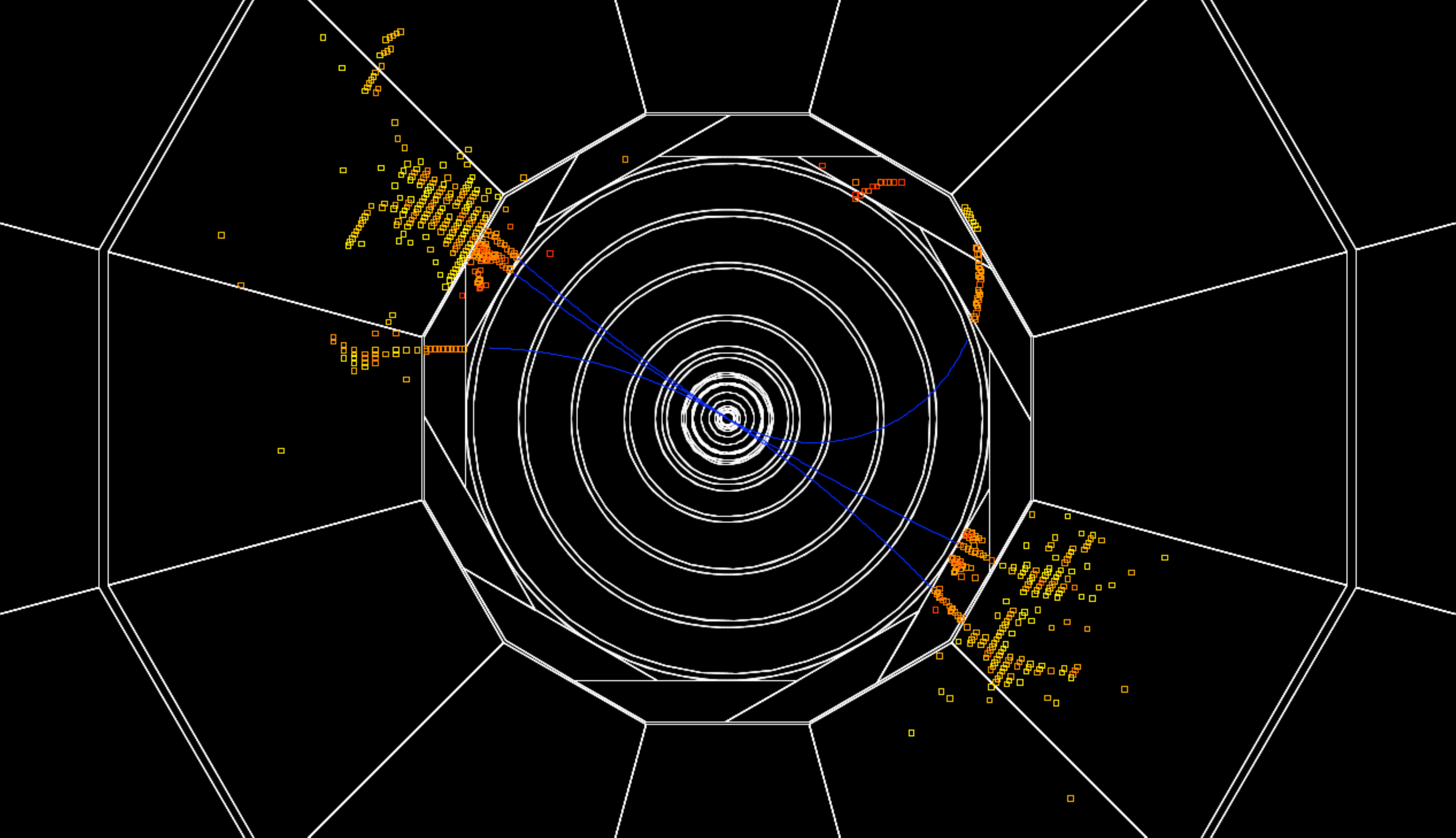}      
    \caption{\small  Wired display of a $h\rightarrow \tau^+\tau^-$ event in the mcdrcal01 detector. }   
    \label{fig:h-tautau-2} 
\end{center}    
\end{figure} 

\subsection{Software Environment}

We used and extended the ALCPG\footnote{American Linear Collider Physics Group} \cite{alcpg} software suite.
Using this software suite enables us to utilize existing standard reconstruction software modules for digitization, 
cluster algorithms, hit manipulation, tracking etc. that are part of the software package. 

The  ALCPG software suite consists of:

\begin{itemize} 
\item SLIC\footnote{Simulator for the Linear Collider}, to simulate the detector response. SLIC is a full simulation 
package that uses the Geant4 Monte Carlo toolkit~\cite{Geant4} to simulate the
passage of particles through the detector.
The SLIC software package uses LCDD\footnote{Linear Collider Detector Description} \cite{lcdd} 
for its geometry input. 
LCDD itself is an extension of GDML\footnote{Geometry Description Markup Language} \cite{gdml}.
LCDD  makes it easy to quickly implement new detector concepts which is especially useful in the early stages of developing and 
optimizing a detector concept .
\item lcsim.org \cite{lcsim}, is a reconstruction and analysis package for simulation studies for the
international linear collider. It is entirely developed in Java for ease of development 
and cross-platform portability. 

\item JAS3\footnote{Java Analysis Studio}\cite{jas3} is a general purpose, open-source data analysis framework. The following
features are provided in the form of plug ins:
\begin{itemize}
 \item LCIO Event Browser \cite{lcio}.
 \item WIRED 4 \cite{wired} is an extensible experiment independent event display.
 \item AIDA\footnote{Abstract Interfaces for Data Analysis} compliant analysis system. It provides tools for 
       plotting of 1d, 2d and 3d histograms, XY plots, scatterplots etc. and fitting (binned or unbinned) using
       an extensible set of optimizers including Minuit.
\end{itemize}
\end{itemize}  


\subsection{Background Rejection Techniques}

The fact that much of the background is soft and out of time gives us two handles on the design 
of an experiment that can cope with the high levels of background.  Timing is especially 
powerful.  The local gate $t = 0$ is defined as the time when a relativistic particle emerging from the interaction 
point arrives at the detector.  Therefore a very tight cut can be made, still preserving the bulk of the tracks of interest. 
A $3$ ns cut rejects two orders of magnitude of the overall background and about four orders of magnitude of 
neutron background.

\begin{figure}[htb] 
  \begin{center}
     \includegraphics[clip=true, scale=0.5]{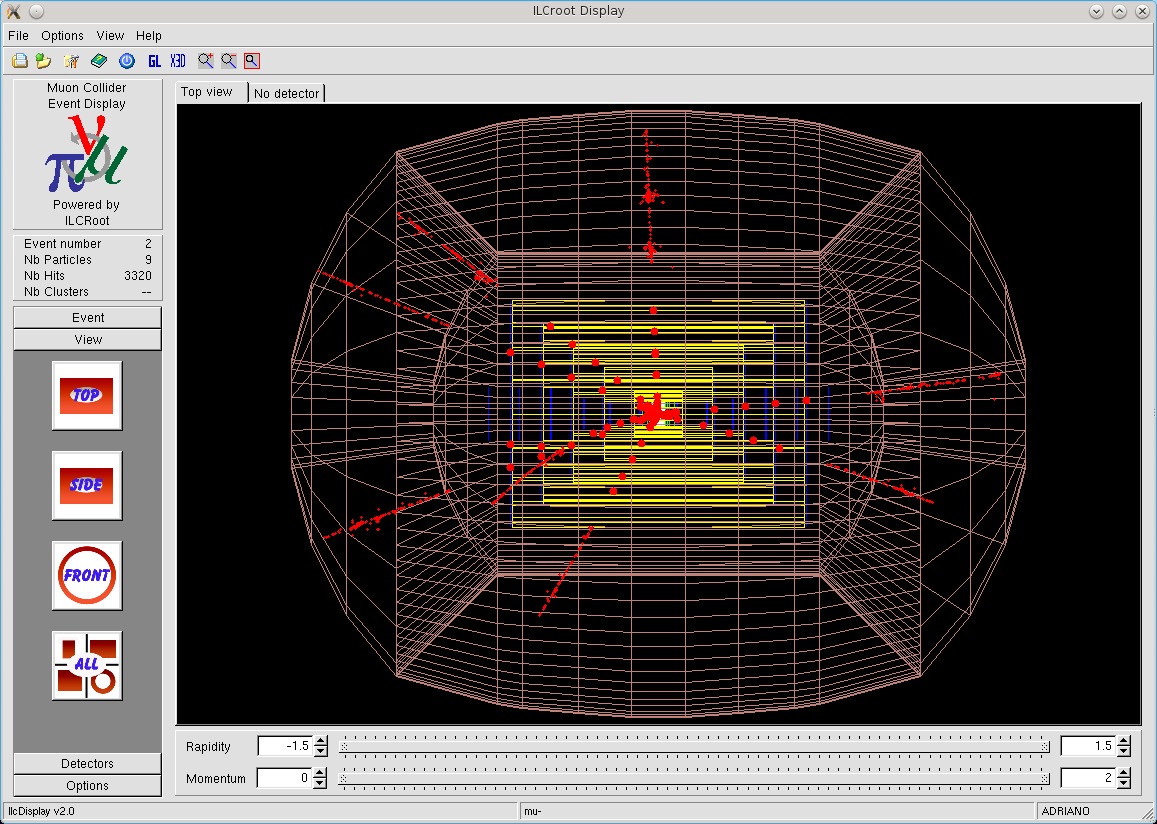}      
    \caption{\small 10 muons from IP displayed in the ILCroot detector developed for  these studies. }   
    \label{fig:10muons} 
\end{center}    
\end{figure} 

A study of timing for hits produced in vertex (VXD) and tracker silicon detectors 
by $0.75\times0.75$ TeV Muon Collider background particles and IP muons was done recently and reported in \cite{Teren}. The ILCroot simulation framework \cite{Gatto} 
was utilized.   
The layout of the VXD and Tracker  is based on an evolution of SiD and SiLC trackers in ILC and developed by the ILCroot group (see detail in \cite{Anna}).
In the analysis the time of flight (TOF) of hits given relative to the bunch crossing time 
was recalculated relative to $T0$, the time of flight for a photon from 
interaction point (IP) ($x=0$, $y=0$, $z=0$) to the detector plane. 
Detector electronics would likely digitize and time stamp hits within a 
larger ($\approx 10$ ns) window to allow for fitting of slow heavy particles 
using TOF as a fitting parameter. The implementation of such time cuts can reduce the occupancy of 
the readout hits in VXD and Tracker layers to the level acceptable for 
efficient tracking of physics tracks as was shown in \cite{Anna}.

\begin{figure}[htb] 
  \begin{center}
    \includegraphics[clip=true, scale=0.75]{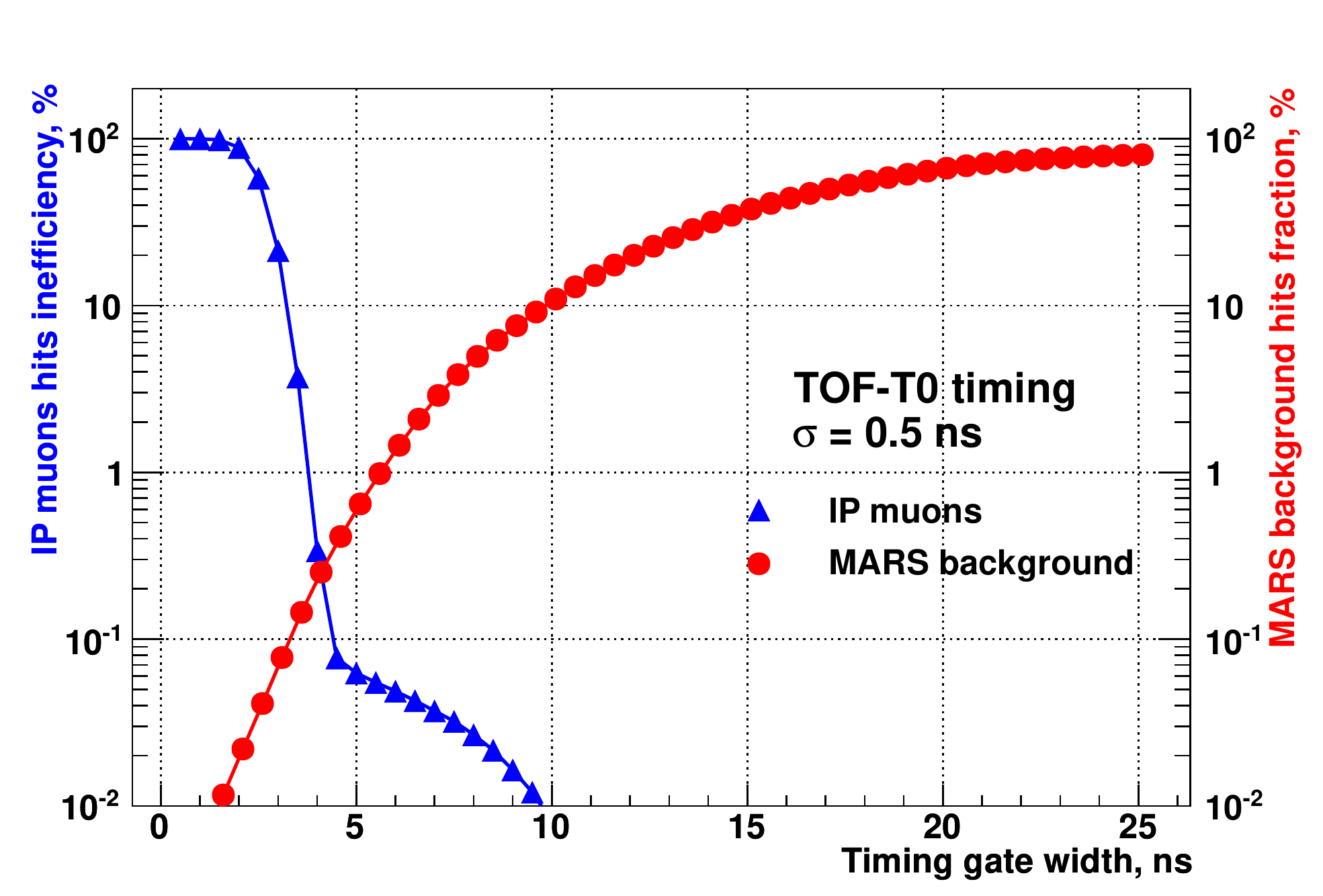}
    \caption{\small  Inefficiency of IP muon hits and fraction of 1.5 TeV collider MARS background particles hits in the VXD and Tracker Si detectors versus width of timing gate at the hit time resolution $\sigma =0.5$ ns.}   
    \label{fig:timing} 
\end{center}    
\end{figure}

Figure \ref{fig:timing} shows IP muon hit inefficiency and fraction of hits from background 
particles versus the timing gate width at 0.5 ns hit resolution time. As we 
can see, a timing gate width of 4 ns can provide  a factor of 300-500 
background rejection keeping efficiency of hits from IP muons higher than 99\%.

\begin{table}
    \begin{center}
    \begin{tabular}{||c|p{2cm}|p{2cm}|p{2cm}|p{2cm}||p{2cm}|p{2cm}|}
    \hline
    \multicolumn{4}{|c|}{Calorimeter background Energy} & \multicolumn{2}{c|}{Tracker Background hits} \\ \hline
    Type                          & Energy before cuts (TeV) & Energy after 2 ns cut (GeV) & Rejection (2 ns cut)  & Radius (cm)             & Rejection (1 ns cut) \\ \hline
    EM               & 170                      & 404                     & $2.4\times 10^{-3} $                & 20                      & $1.2 \times 10^{-3}$              \\
    Muons                         & 185                      & 47                      & $0.25\times 10^{-3}$                & 46                      & $0.8\times 10^{-3}$                \\
    Mesons                        & 7                        & 51                      & $7.5\times 10^{-3}$                 & 72                      & $1.1\times 10^{-3}$                  \\
    Baryons                       & 178                      & 386                     & $2.1\times 10^{-3}$                 & 97                      & $0.6\times 10^{-3}$      \\ \hline
    \end{tabular}
    \label{tab:bkd_rej}
    \caption{Rejection of beam background calorimeter energy and tracker hits for a 1.5 TeV Muon Collider 
    with timing windows with respect to time of flight from the primary vertex of 2 and 1 ns respectively.}
    \end{center}

\end{table}

Timing is also crucial for background rejection in the  calorimeter.  
A calorimeter design studied by R. Raja~\cite{raja1} that combines fast timing with the reconstruction ability of pixelated 
calorimeters being studied for particle flow.  In this design a pixelated  imaging sampling calorimeter with 
$200$ $\mu$m square cells and a 2 ns  ``traveling trigger" gate 
referenced to the time of flight with respect to the 
beam crossing is used to reject out-of-time hits.  
This sort of calorimeter can also implement 
compensation by recognizing  hadronic interaction 
vertices and using the number of such 
vertices to correct the energy.  
Initial estimates of the resolution of such a compensated calorimeter 
is $60\%/\sqrt{E}$. In contrast to relativistic tracks and electromagnetic showers, hadronic showers
can take significant time to develop.
Initial studies of a dual readout total absorption 
calorimeter for the Muon Collider also show that resolution lost to a fast time gate can be regained 
by utilizing a dual readout correction. A summary of the tracking and pixelated calorimetry background rejection 
factors for a 1.5 TeV collider are shown in Table~\ref{tab:bkd_rej}.

We have learned that tracking is feasible in a Muon Collider detector.
Calorimetery is more challenging, but progress is being made on calorimeter concepts
that appear to meet the physics needs.



\newpage

\section{Detector Simulation Studies of Higgs Physics}
	
	We will presently utilize the 
detector model described in the previous section to discuss $s$-channel resonant Higgs boson production and standard model backgrounds at the Muon Collider Higgs Factory.  We assume that we are operating at center-of-mass energy $\sqrt{s} = M_h\approx 125 \gev$ with a beam energy resolution of $\delta E \sim 3.54 \mev$ ($R=0.004\%$).  PYTHIA-generated standard model Higgs boson and background events were used at the generator level to identify and evaluate important channels for discovery and measurement of the Higgs mass, width, and branching ratios.\cite{Conway:2013lca}. 
	
	The $h\rightarrow b\bar{b}$ and $h\rightarrow WW^*$ channels are the most useful for locating the Higgs peak (see Section 2). With an integrated luminosity of $\sim 4.2\ fb^{-1}$ we can measure a $125 \gev$ Higgs boson mass accurately to within $\sim 0.06 \mev $, and the total width to within $\sim 0.18 \mev$. These results demonstrate the value of the large,
resonant Higgs cross-section and the narrow beam energy resolution achievable at a muon collider.

\subsection{Including Backgrounds}

The most significant background for $s$-channel resonance Higgs production at a muon collider is the production of $Z$-bosons. The Higgs cross section, smeared by a $3.54 \mev$ beam is $31.5 \pb$ but first-order initial state radiation corrections bring the effective cross section to $16.7\pb$. The cross section of the $Z$ background is $376 \pb$, but $\sim 20\%$ of these $Z$ decays into pairs of neutrinos and a photon, bringing the cross section to $301.4 \pb$.  This cross section remains essentially flat in the region around the Higgs peak and will be treated as such in this report. 

\begin{figure}[h]
	\centering
	\includegraphics[width=0.9\textwidth]{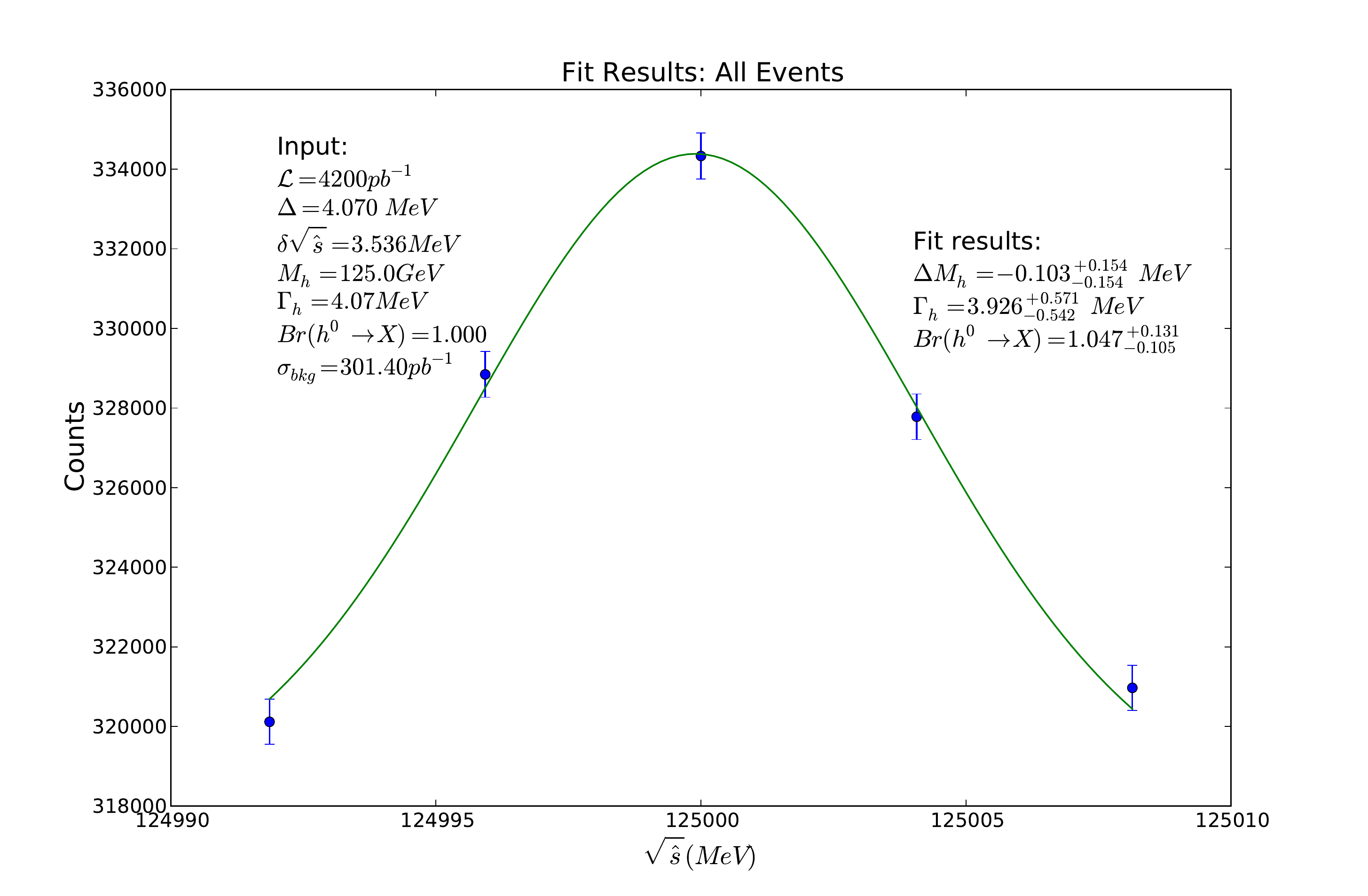}
	\caption{Simulated event counts for a scan across a $125.0 \gev$ Higgs peak with a $3.54 \mev$ wide Gaussian beam spread, counting all events except for $Z^0\rightarrow \nu_{\ell}\bar{\nu_{\ell}}$ decays. Data is taken in a range of $\pm 8.14 \mev$ centered on the Higgs mass in bins separated by the Higgs FWHM (Full width at half maximum) of $4.07 \mev$. Total integrated luminosity is $4.2 \fbi$. Event counts are calculated as Poisson-distributed random variables and the data is fit to a Breit-Wigner convoluted with a Gaussian peak plus linear background. Fitted values of the free parameters are in Table~\ref{table:m-g-meas}.}
\label{data-fit-total-raw}
\end{figure}

Figure~\ref{data-fit-total-raw} shows simulated data of a scan across a $125.0 \gev$  Higgs peak counting all events except for $Z^0\rightarrow \nu_{\ell}\bar{\nu_{\ell}}$. This is fit to a Breit-Wigner distribution, convoluted with a Gaussian with three free parameters, $\Gamma_h$, $M_h$ and Br$(h\rightarrow X)$. The fixed parameters are the background cross section, $\sigma(Z^0\rightarrow X)$, the beam width $\sigma_{beam}$ and the total integrated luminosity $\mathcal{L}$. The fit gives a width of $4.3\pm 0.6 \mev$, an error in the mass measurement of $0.46\pm0.16 \mev$ and a branching ratio of $0.96\pm0.10$. 

\begin{figure}[h]
        \centering
        \includegraphics[width=\textwidth]{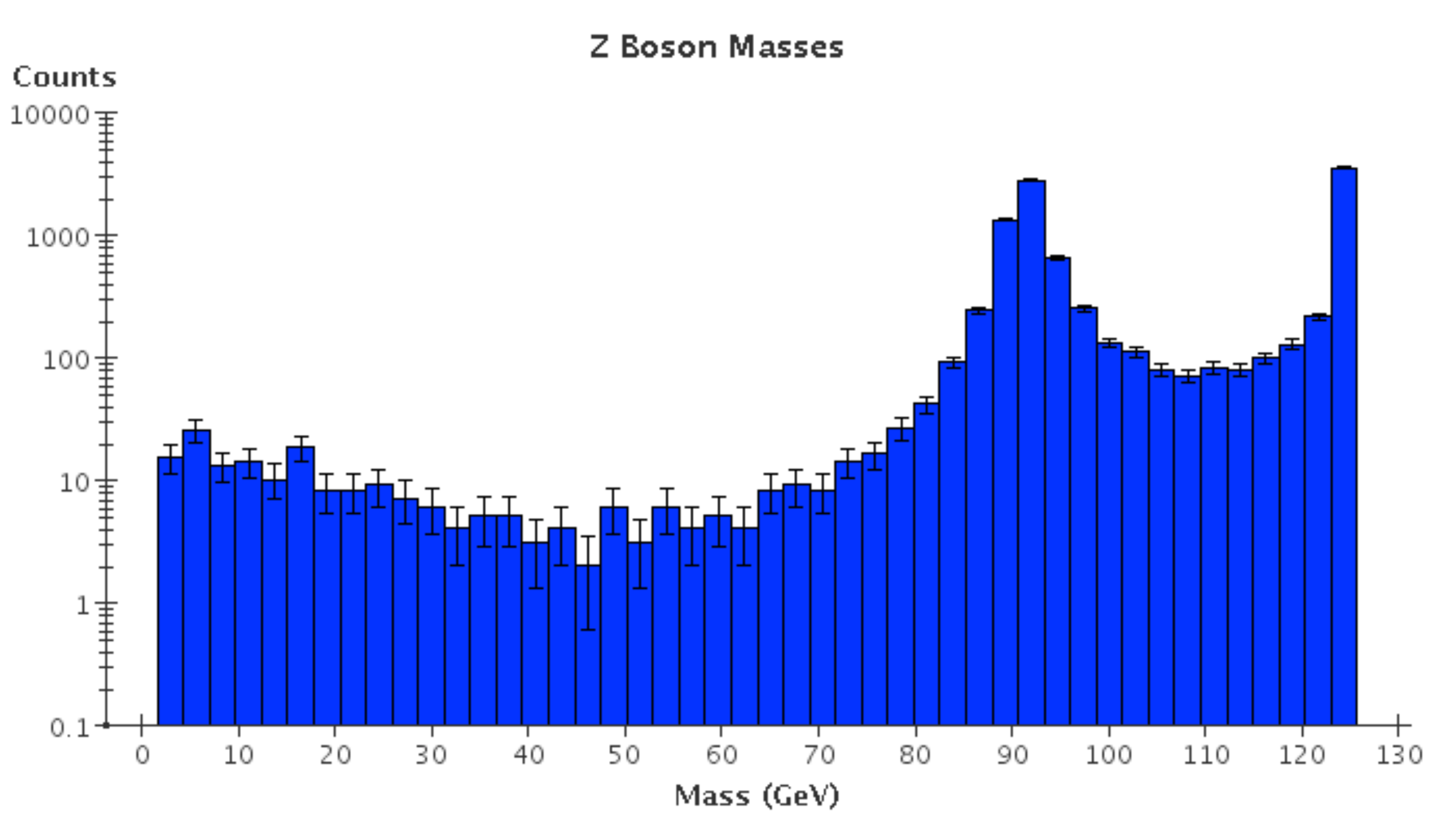}
	\caption[Zboson]{$Z$ boson masses in 10,000 PYTHIA-simulated $\mu^+\mu^-\rightarrow Z$ events at $\sqrt{s}=125.0 \gev$. The low-mass region is dominated by the Drell-Yan process. There is a peak around the $Z$ mass where intial-state Bremsstrahlung radiation allows the creation of an on-shell Z. The third region of interest is the peak at $125 \gev$, the center of mass energy. This represents a process with no initial state radiation where the off-shell Z's produced are indistinguishable from the Higgs.}
\label{z-mass-plot}
\end{figure}

\begin{figure}[ht!]
	\centering
	\subfigure[Irreducible background: $\mu^+\mu^-\rightarrow Z/\gamma^*$ with $M_{Z^*} = \sqrt{s}$.]{
		\includegraphics[width=0.4\textwidth]{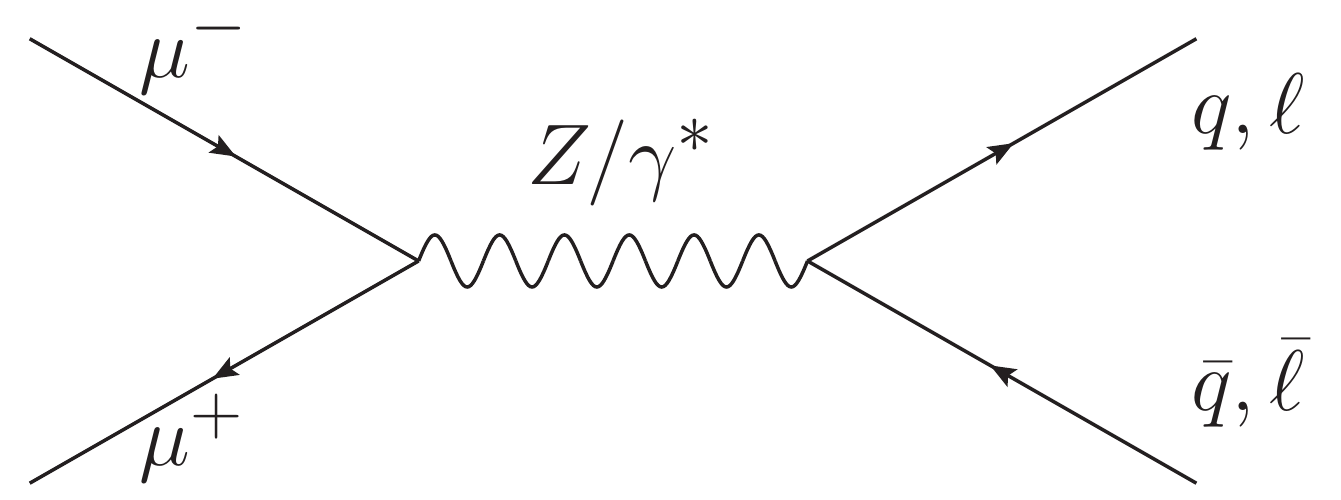}
        \label{fig:mumu-zstar-ql}}
	\qquad
	\subfigure[Reducible background: $\mu^+\mu^-\rightarrow Z^0,\gamma$ with $M_{Z^0} < m_{h}$.]{
		\includegraphics[width=0.4\textwidth]{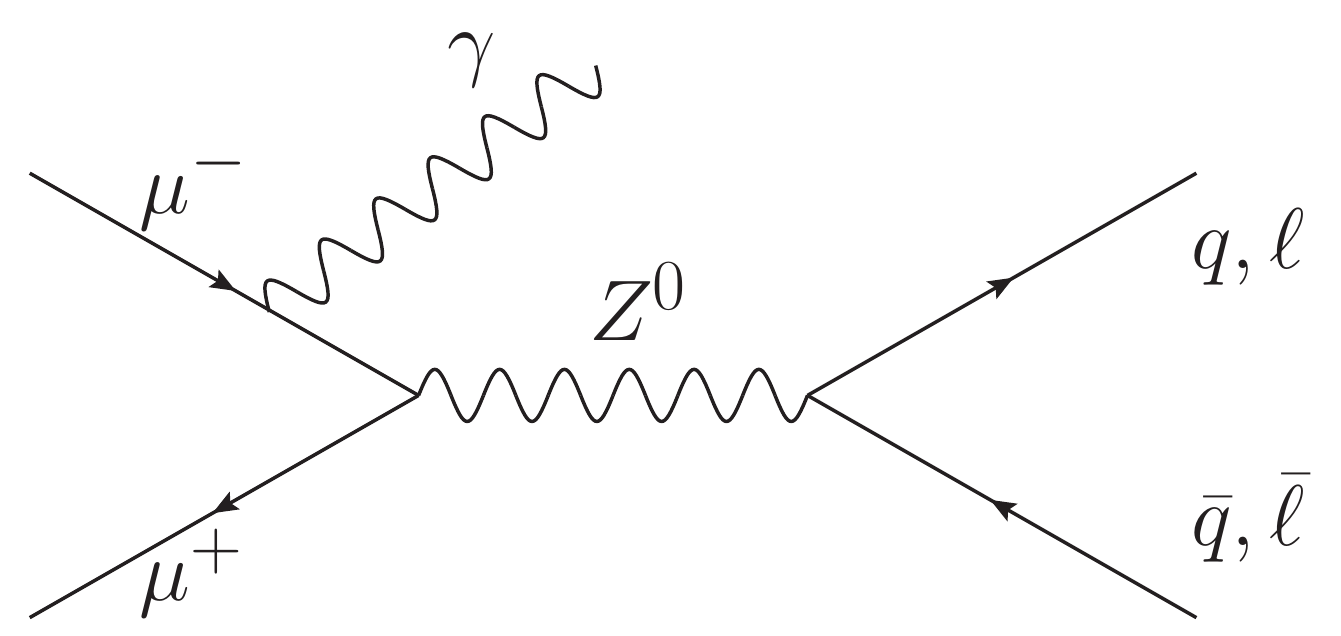}
        \label{fig:mumu-zg-ql}}
	\caption{Standard Model backgrounds at a $\mu^+\mu^-$ collider operating at $\sqrt{s}=126 \gev$}
\label{sm-bkg-feyn}
\end{figure}

	Fortunately, this background is reducible. The $s$-channel resonant production of Higgs bosons only happens within a few $\mev$ of the peak. However, $Z$ bosons are produced in several different processes with a wide range of masses, as seen in Figure~\ref{z-mass-plot}. At an $s$-channel Higgs factory muon collider, $Z$ bosons are primarily produced as real, on-shell bosons along with an intial state photon that makes up the difference in energy between the Higgs s-channel and the $Z$ mass (Fig.~\ref{fig:mumu-zg-ql}). There is also a small number of very low mass $Z$ bosons produced in a Drell-Yan process. The only events that are theoretically indistinguishable from Higgs events are those where a virtual $Z$ is produced at the center of mass energy and decays into a channel shared with the Higgs (Fig.~\ref{fig:mumu-zstar-ql}).

\begin{figure}[h]
	\includegraphics[width=0.9\textwidth]{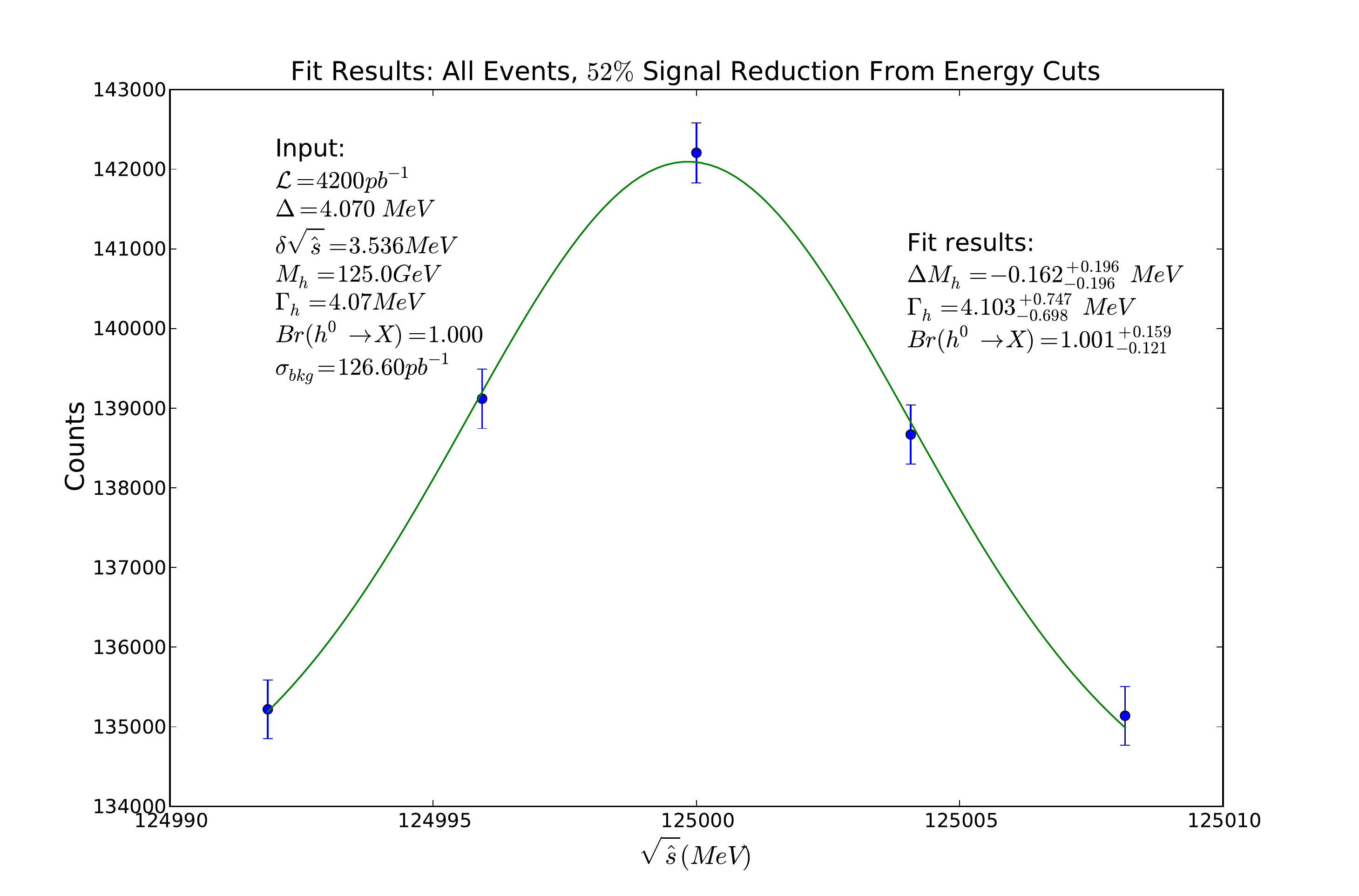}
	\caption{Simulated event counts for a scan across a $125.0 \gev$ Higgs peak with a $3.54 \mev$ wide Gaussian beam spread, counting all events with a total energy of at least $98.0 \gev$ visible to the detector. Data is taken in a $8.14 \mev$ range centered on the Higgs mass in bins separated by the beam width of $4.2 \mev$. Event counts are calculated as Poisson-distributed random variables and the data is fit to a Gaussian peak plus linear background. The fit width is $4.4\pm0.8 \mev$ and the error in the mass measurement is $0.1\pm0.2 \mev$.}
\label{data-fit-total-cut}
\end{figure}

Before looking into how the kinematics of these events might differ from Higgs events, the simple thing to do is a cut on the total energy potentially visible to the detector. This is accomplished by summing the energies of all final state particles which pass a $\cos{\theta} < 0.94$ cut and finding the energy cut which maximizes $S/\sqrt{B}$. The $\cos{\theta}$ cut is effective because most of the high-energy initial state radiation is colinear with the beam. We use a cut of $E_{total} > 98.0 \gev $, which selects 79.2\% of the Higgs signal events and 41.9\% of the $Z$ background. This results in an effective Higgs cross section of $22.4 \pb$ and a background of $126.4 \pb$.. Figure~\ref{data-fit-total-cut} shows simulated data using these results, with a fitted width of $5.57\pm 1.33 \mev$ and an error in the mass measurement of $-0.02\pm0.14 \mev$. This simple cut has already proven to be a marginal improvement but there is much more that can be done by focusing on individual decay channels.

\begin{table}
	\begin{center}
		\begin{tabular}{|c|c|c|c|}
			\hline
			\multirow{2}{*}{Decay Mode} & Bkgr. & \multicolumn{2}{|c|}{$h$} \\ \cline{2-4}
			 & $\sigma$ (pb) & BR & $\sigma$ (pb) \\ 
			\hline
			$u\bar{u}$,$d\bar{d}$,$s\bar{s}$ & 56.0 & 0.0003 & 0.01 \\ \hline
			$c\bar{c}$      & 20.3 & 0.029 & 1.1 \\ \hline
			$b\bar{b}$      & 57.2 & 0.577 & 9.6 \\ \hline
			$e^+e^-$        & 9.2 & --- & --- \\ \hline
			$\mu^+\mu^-$    & 4.8 & --- & --- \\ \hline
			$\tau^+\tau^-$  & 9.2 & 0.063 & 2.4 \\ \hline
			$\nu_{\ell}\bar{\nu_{\ell}}$    & 75.2  & --- & --- \\ \hline
			$gg$                    & ---   & 0.086 & 3.2 \\ \hline
			$\gamma\gamma$  & 27.6   & 0.002 & 0.07 \\ \hline
			$WW^*$                  & 0.001   & 0.215 & 8.0 \\ \hline
			$Z^0Z^0$                & ---   & 0.026 & 0.97 \\
			\hline \hline
			Total:  & 259.6 & 1.0 & 25.5 \\ \hline
		\end{tabular}
	\end{center}
	\caption{Branching fractions and effective cross sections for Standard Model decay modes of Higgs and $Z$ bosons. Higgs cross sections are calculated as the peak value of the Higgs peak Breit-Wigner convoluted with a Gaussian of width $3.54 \mev$ to simulate the effect of beam smearing. The inclusion of initial state radiation effects and full one loop corrections will further reduce the cross sections for Higgs production by a factor of 0.53; resulting in a total Higgs cross section of $13.6 {\rm ~pb}$. Background cross sections are taken from PYTHIA 6.4 event generation output.\label{bfs-xsects}}
\end{table}

\subsubsection{$h \rightarrow b\bar{b}$}

	Table~\ref{bfs-xsects} compares the branching ratios and cross sections of the $Z$ background with the Higgs signal. The largest Higgs decay channel is $h\rightarrow b\bar{b}$, which makes up 58\% of Higgs decays at this mass, a branching fraction proportionally large to $Br(Z^0\rightarrow b\bar{b}) = 15.2\%$. We assume a b-tagging efficiency and purity of 1, so the cross sections for the decays are 16.5 and 57.2 pb, respectively. The fitted values for the mass, width and branching ratio of the Higgs using b-tagging are shown in Table~\ref{table:m-g-meas}.

	In both signal and background the visible energy spectrum is very similar to the spectrum of the combined channels, so the same total energy cut of $E_{tot}> 98.0 \gev$ maximizes $S/\sqrt{B}$. Cuts on the event shape, the magnitude of the thrust and major axis, can further enhance the signal. The event shape is calculated by finding the axis which maximizes the sum of all particle momenta projected onto a single axis, called the `thrust axis'. This is then repeated for an axis perpendicular to the first and then a third orthogonal to both. The thrust is the normalized sum of the projection of all particle momenta against the thrust axis and the major axis value is the normalized sum of the projections against the secondary axis. Because the Higgs is never created in events with significant beamstrahlung it is always produced with low momentum. $Z$ bosons produced with mass lower than the beam center-of-mass energy are `boosted' by the beamstrahlung photon. This boost lowers the thrust and raises the major axis values, so it is a useful indicator for channels with particular event shape profiles.
	
\begin{figure}
	\includegraphics[width=\textwidth]{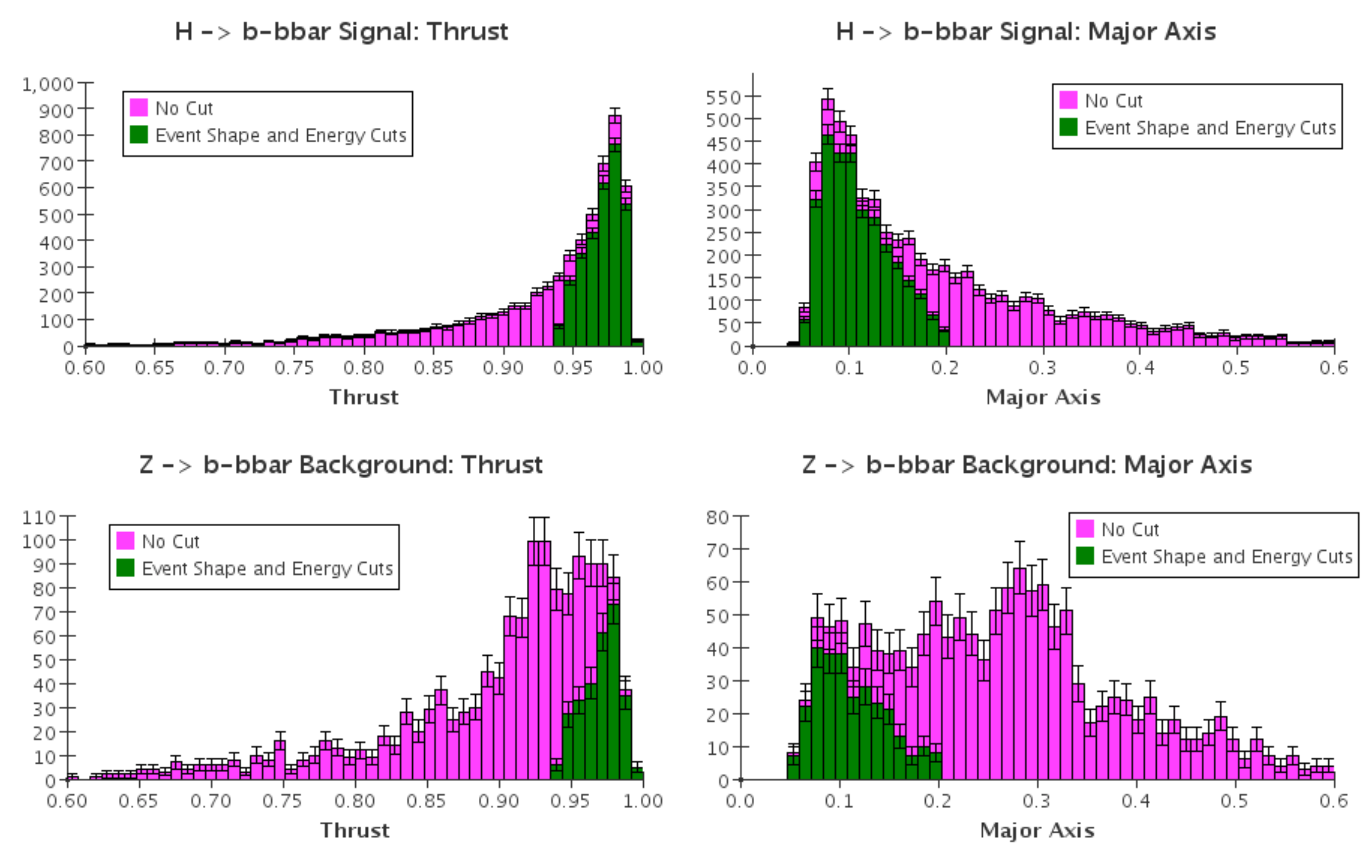}
	\caption{Effects of event shape and energy cuts on Higgs $b\bar{b}$ signal and background. Cuts were made by selecting events with total energy $E_{tot} > 98.0 \gev$ visible to the detector, thrust between 0.94 and 1.0 and major axis between 0.0 and 0.2. The signal is reduced to 52\% and the background to 15\%.}
\label{bbbar-en-thrust-cuts}
\end{figure}

\begin{figure}
	\includegraphics[width=\textwidth]{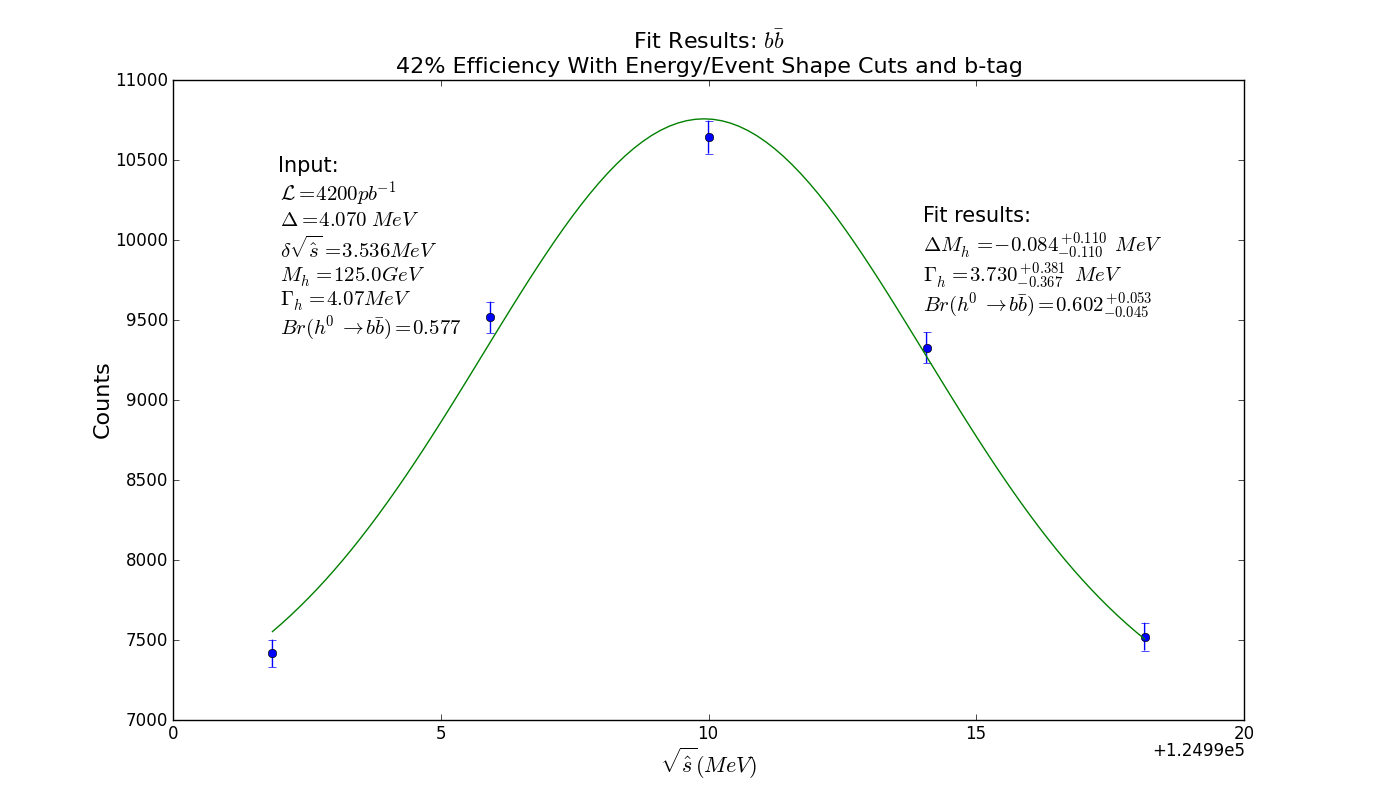}
	\caption{Simulated event counts for a scan across a $125.0 \gev$ Higgs peak with a $4.2 \mev$ wide Gaussian beam spread, counting $X\rightarrow b\bar{b}$ events with a total energy of at least $98.0 \gev$ visible to the detector and cutting on event shape parameters. Data is taken in a $60 \mev$ range centered on the Higgs mass in bins separated by the beam width of $4.2 \mev$. Event counts are calculated as Poisson-distributed random variables and the data is fit to a Breit-Wigner convoluted with a Gaussian plus linear background. The fit width is $4.78\pm0.48 \mev$, the error in the mass measurement is $0.01\pm0.05 \mev$ and the branching ratio is measured at $0.271\pm0.001$. Total luminosity is $1000 \pbi$, or $71.4 \pbi$ per point.}
\label{data-fit-bbbar-cut}
\end{figure}

Figure~\ref{bbbar-en-thrust-cuts} shows the signal and background thrust and major axes before and after cutting on the total energy and event shape values. The cuts were made by selecting events with $E_{tot} > 98.0 \gev$, thrust values between 0.94 and 1.0 and major axis values between 0.0 and 0.20. We continue to assume perfect b-tagging. These cuts reduce the $b\bar{b}$ signal by 52\% and the background by 15\%, bringing the effective cross sections to 8.64 and 8.45 pb respectively. This improves the $S/\sqrt{B}$  over simple energy cuts or b-tagging alone. Figure~\ref{data-fit-bbbar-cut} shows a simulated scan of the Higgs peak with a fit to a Breit-Wigner convoluted with a Gaussian.

\subsubsection{$h\rightarrow WW^*$}

There are several channels with very little physics background that are of importance, despite their smaller cross sections. One of these is the $h\rightarrow WW^*$ decay mode, with a branching fraction of 0.226 (cross section 6.39 pb) and no real background from the corresponding $Z$ decays. The $W$ boson decays into a charged lepton and corresponding neutrino 32.4\% of the time, with effectively equal rates for each type of lepton. The majority of the remaining branching fraction is the decay into pairs of light quarks. While it is certainly possible to reconstruct $W$ bosons from four-jet events, in this report we focus on the decays with missing energy in the form of neutrinos since they can be identified by the presence of one or two isolated leptons and missing energy and are the most common. Further study will be required for a detailed analysis of the four-jet case. Since the $W$ boson decays into a lepton and neutrino 32.4\% of the time and we require at least one such decay between a pair of $W$'s, these make up 54.3\% of $WW^*$ events. Thus the theoretical cross section is 6.39 pb with virtually no background.

Because the detector will have a non-sensitive cone, there will be a small amount of `fake' background, eg.~when the photon in the decay $\mu^+\mu^-\rightarrow Z^0+\gamma\rightarrow\ell^++\ell^-$ boosts the two leptons and disappears into the cone as missing energy. 
It is difficult to estimate the true background from processes such as these, but given the low branching ratios of $Z^0$ to lepton pairs and the kinematic and geometric constraints for `fake' background, it is safe to assume that the background will be fairly low in this channel. Therefore we use a cross-section of 0.051 pb. 

\subsubsection{$h\rightarrow \tau^+\tau^-$}

The $\tau^+\tau^-$ channel is dominated by the background, but the Higgs branching ratio of 0.071 is not insignificant. The $Z^0\rightarrow \tau^+\tau^-$ process has a branching ratio of $0.034$, giving it an effective cross section of $12.8$ pb, compared to the $2.01$ pb cross section for the Higgs. However, the boost given to the lower mass $Z$ bosons means the background can be further distinguished using total energy and event shape parameters. 

\begin{figure}
	\includegraphics[width=\textwidth]{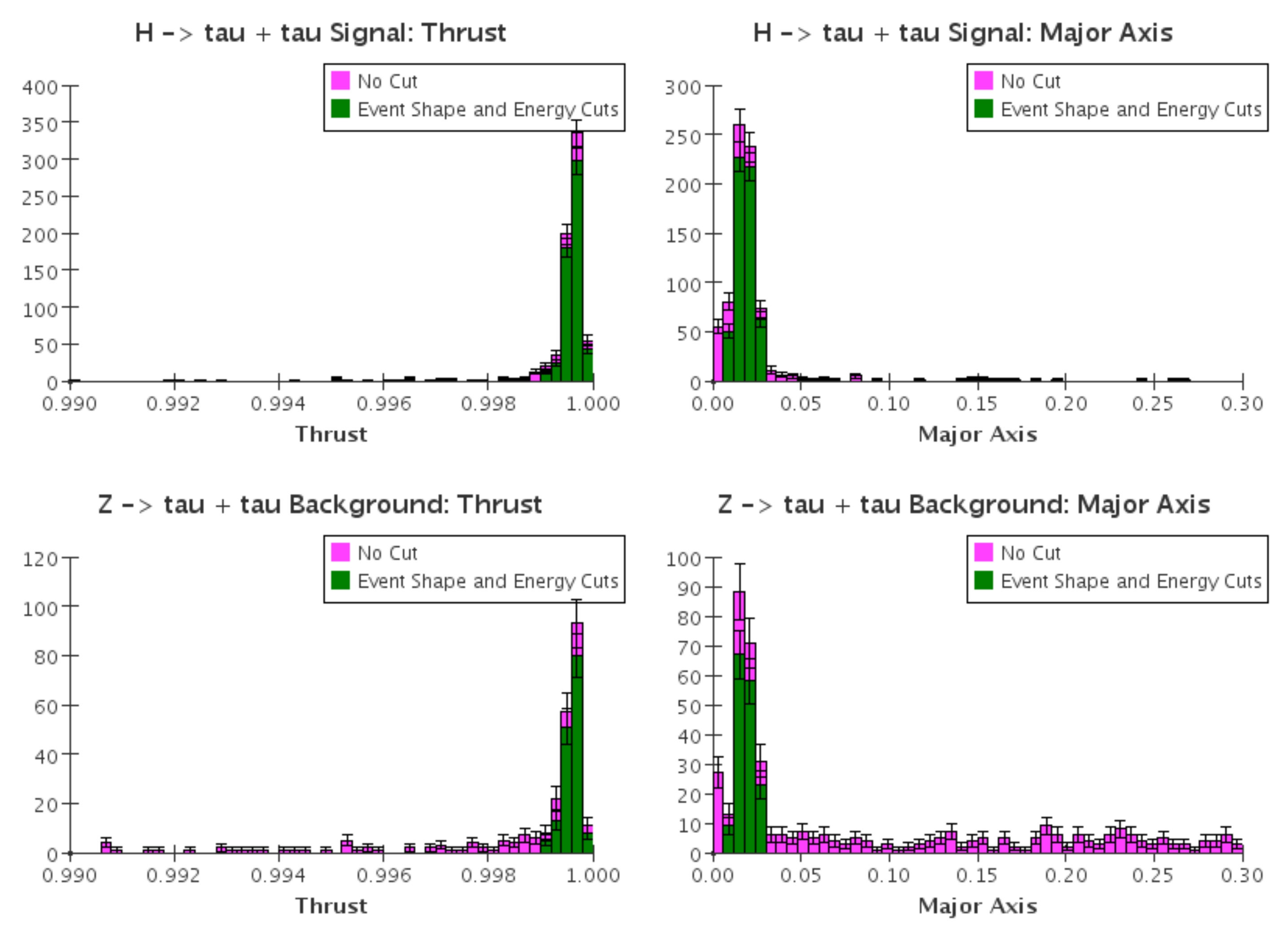}
	\caption{Effects of event shape and energy cuts on Higgs $\tau^+\tau^-$ signal and background. Cuts were made by selecting events with total energy $E_{tot} > 60.0 \gev$ visible to the detector, thrust between 0.999 and 1.0 and major axis between 0.07 and 0.032. The signal is reduced to 78\% and the background to 39\%.}
\label{tautau-en-thrust-cuts}
\end{figure}

The $\tau$ is a short-lived particle and every $\tau$ decay channel involves the production of a $\tau$ neutrino. This makes the total visible energy less useful as a cut parameter than it was for $b\bar{b}$, since there are random amounts of missing energy. We require at least $60.0 \gev$ to be visible because background dominates below this value due to boosted $Z$'s. Event shape parameters, however, are very useful here since $\tau$ decays typically do not create a widespread shower. We require the thrust to be between 0.999 and 1.0 and the major axis to be between 0.007 and 0.03. This cut reduces the signal to 78\% of its original value and the background to 39\%, bringing the Higgs cross section to 1.58 pb and the background to 4.97 pb, as seen in Figure~\ref{tautau-en-thrust-cuts}. The cut is specific enough that it is not necessary to assume anything else about the events, such as a perfect $\tau^+\tau^-$ tag. Fewer than 0.2\% of the Higgs decays that pass the cut are not $\tau^+\tau^-$ events and only 6.4\% of the background events that pass are misidentified. The effective background cross section above is calculated from all the events which pass the cut. 

\subsubsection{$h \rightarrow c\bar c$}

With an integrated luminosity of $\sim1 \fbi$ one expects $\sim 23,000$ produced Higgs bosons,  implying  $\sim 800$ $h\rightarrow c\bar c$ decays and
$\sim 13,000$ $h\rightarrow  b \bar b$  decays. 

To obtain a sensitive $ c\bar c$
branching fraction implies the need to reject the
$b\bar b$ background by a factor of 20 or more. Additionally, there is a
long tail of $c\bar c$ and $b \bar b$  coming from $Z$ decays. 
This produces a background of 19 pb under the
$h$ peak \cite{Eichten:2013ucla}, and it therefore generates an additional
$\sim 19,000$ events. 

\begin{figure}[htb]
  \begin{center}
   \includegraphics[width=100mm]{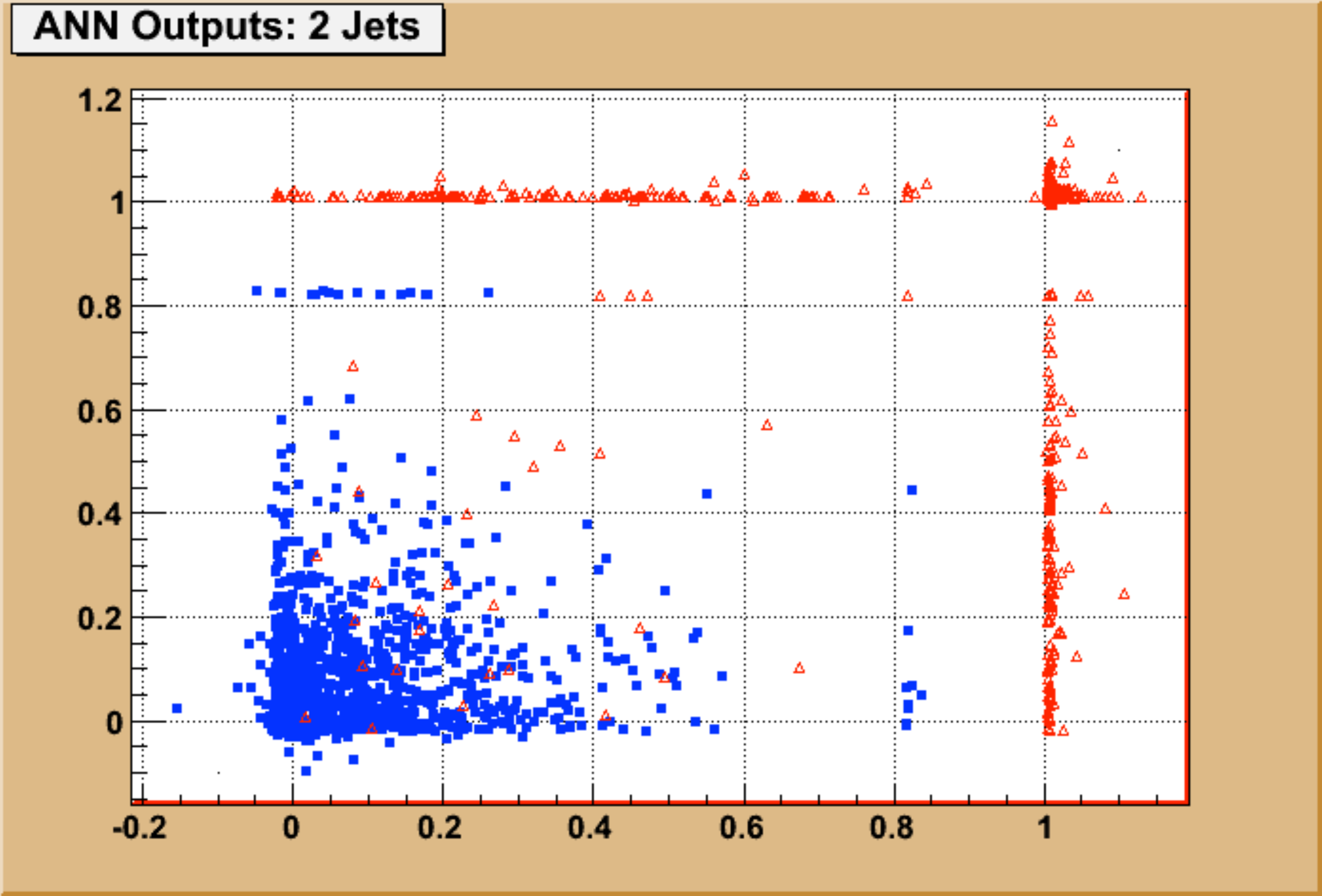}
   \caption{The two-jet neural net output plane for $\mu^+ \mu^- \rightarrow h \rightarrow 2 {\rm~heavy~ quark~jets}$ events in the detector. 
   True $c$-jets are shown as blue squares, while the true $b$-jets are shown as red triangles.}
   \label{fig:cc_2jets}
 \end{center}
\end{figure}

Observing the $h$ peak in $c \bar c$ decays should still be possible, but the background
from $Z$-decays will be the dominant one after rejection of the $h \rightarrow b \bar b$ 
decays. As shown in Fig.~\ref{fig:cc_2jets}, excellent
separation of charm and bottom jets is also possible for Higgs decays at
the muon collider.   A simple estimate of $S/\sqrt{B}$ implies an
observation of the $h\rightarrow c \bar c$ signal 
will be possible at the $\sim 5$ to $\sim 6$ $\sigma$ level
at the muon collider Higgs factory \cite{Purohit:2013ucla}.

%

\begin{table}
	\centerline{
	\begin{tabular}{|c|l|l|l|l|l|l|l|}
		\hline
		\multirow{2}{*}{Channel} & \multicolumn{3}{|c|}{$\mu^+\mu^-\rightarrow h \rightarrow X$} & \multicolumn{3}{|c|}{$\mu^+\mu^- \rightarrow X$} & \multirow{2}{*}{$S/\sqrt{B}$} \\
		\cline{2-7}
		& Br &\multicolumn{2}{|c|}{$\sigma$ (pb)} & Br & \multicolumn{2}{|c|}{$\sigma$ (pb)} &  \\
		\hline
		 \multirow{2}{*}{Total}
			& \multirow{2}{*}{1.0}	& $\sigma_s$		& 16.7	
			& \multirow{2}{*}{1.0}	& $\sigma_b$		& 376.0	& 55.8	\\
			&&$\sigma_{eff}$ & 13.2 && $\sigma_{eff}$& 126.6	& 76.0	 \\
		\hline
		\multirow{2}{*}{$b\bar{b}$}	
			& \multirow{2}{*}{0.577}	& $\sigma_s$	& 9.6
			& \multirow{2}{*}{0.152}	& $\sigma_b$	& 57.2	& 82.3	 \\
			&&$\sigma_{eff}$ & 4.0	&	& $\sigma_{eff}$& 6.9	& 98.7	 \\
		\hline
		\multirow{2}{*}{$WW^*$}	
			& \multirow{2}{*}{0.215}	& $\sigma_s$	& 3.6	
			& \multirow{2}{*}{5e-4}	& $\sigma_b$	& ---	& ---		 \\
			&&$\sigma_{eff}$ & 1.9	&	& $\sigma_{eff}$& 0.001	& 3,894	 \\
		\hline
		\multirow{2}{*}{$\tau^+\tau^-$}	
			& \multirow{2}{*}{0.063}	& $\sigma_s$	& 1.1	
			& \multirow{2}{*}{0.042}	& $\sigma_b$	& 12.8	& 19.9	 \\
			&&$\sigma_{eff}$ & 0.9	&	& $\sigma_{eff}$& 5.1	& 23.0	 \\
		\hline
		\multirow{2}{*}{$c\bar{c}$}	
			& \multirow{2}{*}{0.029}	& $\sigma_s$	& 0.48	
			& \multirow{2}{*}{0.118}	& $\sigma_b$	& 44.4	& 12.6	 \\
			&&$\sigma_{eff}$ & 0.44	&	& $\sigma_{eff}$& 42.9	& 11.9	 \\
		\hline
		\multirow{2}{*}{$\gamma\gamma$}	
			& \multirow{2}{*}{0.002}	& $\sigma_s$	& 0.03	
			& \multirow{2}{*}{0.13	}	& $\sigma_b$	& 27.6	& 0.37	 \\
			&&$\sigma_{eff}$ & 0.02	&	& $\sigma_{eff}$& 27.6	& 0.6	\\
		\hline
	\end{tabular}
	}
	\caption{Branching fractions, cross sections before and after cuts and $S/\sqrt{B}$ for the channels studied. We assumed an integrated luminosity of $4.2 \fbi$ for determining the $S/\sqrt(B)$ values. Background cross sections were calculated with PYTHIA 6.4 simulations.}
\label{table:m-g-meas}
\end{table}

\begin{table}
	\centerline{
	\begin{tabular}{|c|r|l|l|l|}
		\hline
		\multirow{2}{*}{Channel}	&\multirow{2}{*}{}	& \multirow{2}{*}{$\Gamma_{H\rightarrow X} (MeV)$}	& \multirow{2}{*}{$\Delta M_H (MeV)$}	& \multirow{2}{*}{$Br(h\rightarrow X)$} \\
		&	&	&	&\\
		\hline
		\multirow{2}{*}{Total}	&
				Raw	& $3.9\pm0.6$	& $-0.10\pm0.15$	&$1.05\pm0.13$ \\
			  &	Cut	& $4.1\pm0.7$	& $-0.16\pm0.19$& $1.0\pm0.16$ \\
		\hline
		\multirow{2}{*}{$b\bar{b}$}	&
				Raw	& $4.3\pm0.5$	& $0.1\pm0.1$ & $0.55\pm0.05$ \\
		&		Cut	& $3.7\pm0.4$	& $-0.08\pm0.1$	 & $0.60\pm0.05$ \\
		\hline
		\multirow{2}{*}{$WW^*$}	&
				Raw	& $---$	& $---$ & $---$ \\
			  &	Cut	& $3.9\pm0.2$	& $0.06\pm0.07$ & $0.22\pm0.01$ \\
		\hline
		\multirow{2}{*}{$\tau^+\tau^-$}&
				Raw	& $3.5\pm2.0$	& $0.00\pm0.5$ & $0.07\pm0.05$ \\
		&       Cut	& $4.5\pm1.5$	& $-0.1\pm0.4$  & $0.06\pm0.02$ \\
		\hline
	\end{tabular}
	}
	\caption{Fitted values of Higgs decay width, mass and branching ratio from simulated data. Mass values are the difference between the measured mass and the true mass of $126,000 \mev$. Total integrated luminosity was $4.2 \fbi$, or $840 \pbi$ per data point.}
\label{table:h-fits}
\end{table}

\subsection{Combining Channels}

\begin{table}
	\begin{center}
		\begin{tabular}{|c|l|l|l|}
			\hline
			Channel	& $\delta M_H$ (MeV)	& $\delta \Gamma_H$ (MeV)	& $\delta Br(h\rightarrow X)$ \\ \hline
			$b\bar{b}$	& 0.1	& 0.4	& 0.05	\\ \hline
			$WW^*$		& 0.07	& 0.2	& 0.01	\\ \hline
			Combined	& 0.06	& 0.18	& ---	\\ \hline
		\end{tabular}
	\end{center}
	\caption{Accuracy of fitting parameters for simulated Higgs data using individual and combined channels.\label{table:fitted-vals-boxpl}}
\end{table}

To measure the Higgs mass and total width more precisely, we took advantage of both the $b\bar{b}$ and the $WW*$ channels. We did this by simulating data for both channels and taking their average, weighted by the uncertainty in the fits, with results shown in Table \ref{table:fitted-vals-boxpl}. For example, the formula used for the width was:
\begin{equation}
	\overline{\Gamma}_H = \frac{\delta \Gamma_{b\bar{b}}}{\delta \Gamma_{b\bar{b}} + \delta \Gamma_{WW^*}}\Gamma_{WW^*} + \frac{\delta \Gamma_{WW*}}{\delta \Gamma_{b\bar{b}} + \delta \Gamma_{WW^*}}\Gamma_{b\bar{b}}\label{eq:weight-avg}
\end{equation}
\begin{equation}
	\delta \overline{\Gamma}_H^2 = \frac{\delta \Gamma_{b\bar{b}}^2\delta \Gamma_{WW*}^2} {\delta \Gamma_{b\bar{b}}^2 + \delta \Gamma_{WW*}^2}\label{eq:delta-avg}
\end{equation}

\subsection{Potential to Resolve Nearly Degenerate Higgs Bosons}
\label{sec:degenerate}

The Higgs line-shape scanning process not only provides high precision for the Higgs boson total width, but also high precision for the Higgs mass. Sub MeV level precision can be achieved as show in Table~\ref{tab:acrcy}. This implies that the muon collider will be an ideal machine to break the mass degeneracies between Higgs bosons. In this section, we discuss this potential of the muon collider.

There are theoretical speculations that the LHC Higgs signal may be a combination of two nearly degenerate Higgs-like bosons~\cite{Ferreira:2012nv,Celis:2013rcs,Efrati:2013tia,Gunion:2012gc,Cerdeno:2013cz,Christensen:2013dra,Grossman:2013pt}. This could happen in some models, for example, two Higgs doublet models (2HDM)~\cite{Ferreira:2012nv,Celis:2013rcs,Efrati:2013tia}, 2HDM plus 1 singlet models, as well as next-to-minimal-supersymmetric-model~\cite{Gunion:2012gc,Cerdeno:2013cz,Christensen:2013dra}. A putative $\sim 1 \gev$ level degeneracy could be easily resolved at an early stage of the muon collider program, when  the Higgs mass window is determined, 
as described in Sec.\ref{sec:window}. However, we can go much further, and we discuss presently
the $\sim 1 \mev$ level mass degeneracy resolution 
achievable at a muon collider.

\begin{figure}[htb]
\begin{center}
\includegraphics[width=250pt]{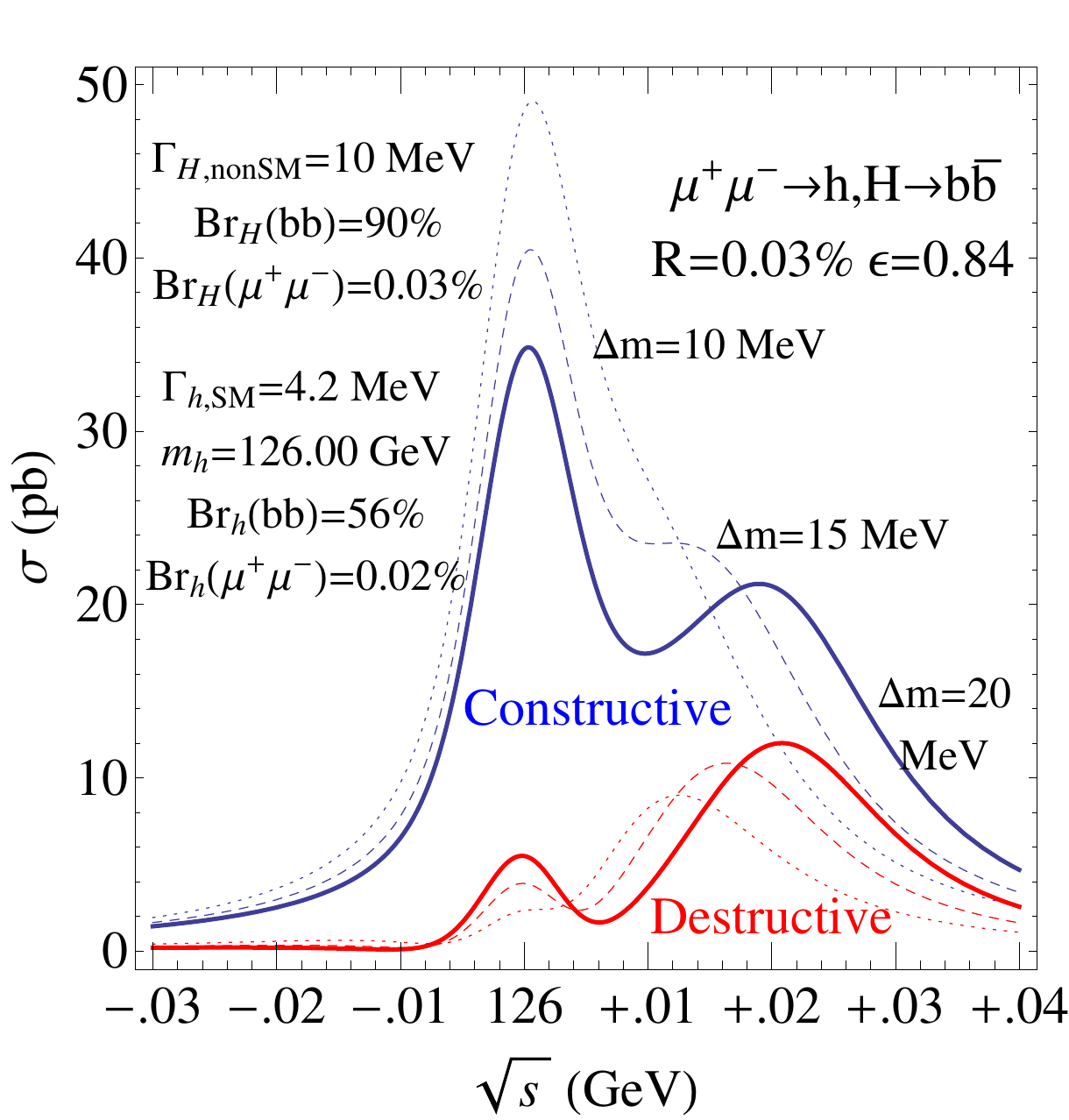}
\caption{Resolving highly degenerate Higgs bosons at the muon collider through scanning. The b-tagging efficiency is assumed to be $60~\%$, and the acceptance $\epsilon$ is thus $0.84$ with at least one b-jet tagged. The solid, dashed and dotted lines represent mass splittings of the Higgs bosons, $20~\mev,~15~\mev$ and $10~\mev$. The blue and red curves represent constructive and destructive interferences, respectively.
}
\label{fig:peaks}
\end{center}
\end{figure}

A naive expectation is that the muon collider could resolve the mass degeneracy to sub MeV level, as it does for the single Higgs boson mass fitting. However, this is way below the beam energy spread and the Higgs boson total width. The latter means the interference effect between these two highly degenerate Higgs bosons has to be taken into account at the muon collider. We demonstrate this resolution in Fig.~\ref{fig:peaks} for $\mu^+\mu^-\rightarrow h,H\rightarrow b\bar b$. We fix the SM Higgs Boson at $126~\gev$ with total width $4.2~\mev$. We set the other Higgs total width at $10~\mev$ and the branching fractions to $b\bar b(\mu^+\mu^-)$ $90\%(0.03\%)$. This corresponds to a non-SM doublet Higgs with $\tan\beta$ around $2$ in Type II 2HDM, as well as (more likely) a larger $\tan\beta$ with a significant mixing with additional singlet. We choose three different masses from $126.01~\gev$ to $126.02~\gev$ and demonstrate both constructive  and destructive interferences.

This mass degeneracy resolution depends on many factors.  The optimal scenario would be both Higgs bosons having the same total width and signal strength. Instead of this optimal scenario, our choice in Fig.~\ref{fig:peaks} has both Higgs bosons with the same order of strength and total width. We can see that shape fitting is necessary to resolve a $\sim 10~\mev$ degeneracy. As a result, we argue the muon collider could a resolve mass degeneracy to the level of these Higgs bosons' total widths.

There are other ways to resolve the mass degeneracy at the muon collider. For example, for 2HDM and related models, the other 
Higgs is expected not to
have suppressed couplings to the vector bosons. One could fit the mass from the $WW^*$ mode to sub MeV level for the SM-like Higgs and fit the mass from  the $b\bar b$ mode to a similar level.
These two fits would have different best fit masses and thus resolve the degeneracy.
This scenario dependent method has the potential of resolving the mass degeneracy to the MeV level.

The muon collider Higgs factory is an ideal place to resolve the mass degeneracy of Higgs bosons. Its resolution is at the level of the Higgs bosons' total widths. This excellent mass degeneracy resolution can also be applied to the future upgrade of the muon collider for the energy frontier, where in many 2HDM and related models the heavier Higgs and CP-odd Higgs are highly degenerate \cite{Eichten:2013ckl}.


\subsection{Testing the CP Property}

While the newly discovered Higgs boson seems to behave a SM-like, the nature of its couplings to fermions is largely unknown. Denote a generic Higgs scalar ($H$) to couple to a pair of fermions ($\psi$) by 
\be
\label{eq:cpv}
H\ \anti\psi (a+ib\gamma_5) \psi.  
\ee
The most ideal means for determining the CP nature of
a Higgs boson at the muon collider is to employ transversely polarized muons \cite{Barger:2001mi,Grzadkowski:2000hm}.  For $\h$ production at a muon collider  with muon coupling
given by the form in Eq.~(\ref{eq:cpv}),
the cross section takes the form
\bea
\overline\sigma_{\h}(\zeta)&=&\overline\sigma_{\h}^0\left(
1+P_L^+P_L^-+P_T^+P_T^-\left[{a^2-b^2\over a^2+b^2}\cos\zeta-{2ab\over a^2+b^2}\sin\zeta \right]
\right)\nonumber\\
&=&\overline\sigma_{\h}^0
\left[1+P_L^+P_L^-+P_T^+P_T^-\cos(2\delta+\zeta)\right]\,,
\eea
where $\overline \sigma_\h^0$ is the unpolarized convoluted cross section,
$\delta\equiv \tan^{-1}(b/ a)$, 
$P_T$ ($P_L$) is the degree of transverse (longitudinal)
polarization, and $\zeta$ is the angle between the $\mu^+$  and $\mu^-$ transverse polarizations.
Only the $\sin\zeta$ term is genuinely CP-violating, but the $\cos\zeta$ term
also provides significant sensitivity to $a/b$. Ideally, one would like to isolate
${a^2-b^2\over a^2+b^2}$  and ${-2ab\over a^2+b^2}$ by
running at fixed angles $\zeta=0,\pi/2,\pi,3\pi/2$ and measuring the
asymmetries. Taking $P_T^+=P_T^-\equiv P_T$ and $P_L^\pm=0$, we form two observables
\bea
{\cal A}_I&\equiv& {\overline\sigma_{\h}(\zeta=0)-\overline\sigma_{\h}(\zeta=\pi) 
\over \overline\sigma_{\h}(\zeta=0)+\overline\sigma_{\h}(\zeta=\pi) }=
P_T^2{a^2-b^2\over a^2+b^2}=P_T^2\cos2\delta\,,\nonumber\\
{\cal A}_{II}&\equiv& {\overline\sigma_{\h}(\zeta=\pi/2)-\overline\sigma_{\h}(\zeta=-\pi/2) 
\over \overline\sigma_{\h}(\zeta=\pi/2)+\overline\sigma_{\h}(\zeta=-\pi/2) }
=-P_T^2{2ab\over a^2+b^2}=-P_T^2\sin2\delta\,.
\nonumber
\eea
If $a^2+b^2$ is already well determined from its overall coupling strength, and the background is known depending on the specific final state, then the fractional error in these asymmetries can be approximated as
${ \delta{\cal A} / {\cal A}}\propto {1 / P_T^2 \sqrt L}$ \cite{Grzadkowski:2000hm}, 
which points to the need for the highest possible transverse polarization, even if some sacrifice in $L$ is required. 

A crude estimate \cite{Grzadkowski:2000hm} showed that with $P_{T}\sim 40\%$ and  $L=0.1\fbi$ delivered on a scalar resonance, a 30\% ($1\sigma$) measurement on $b/ a$ is possible.
In reality the precession of the muon spin in a storage ring makes running at fixed $\zeta$ impossible. Detailed studies will be needed to achieve quantitative conclusions.

\end{document}